\documentclass[twocolumn]{emulateapj}
\usepackage{natbib,lscape}

\shorttitle{On the Limb Darkening of Procyon}
\shortauthors{Aufdenberg et al.}

\begin{document}

\newcommand{\h}{$^{\rm h}$}
\newcommand{\m}{$^{\rm m}$}
\newcommand{\rsun}{$R_{\odot}$}
\newcommand{\msun}{$M_{\odot}$}
\newcommand{\lsun}{$L_{\odot}$}
\newcommand{\teff}{\ensuremath{T_{\rm eff}}} 
\newcommand{\kms}{km s$^{-1}$}
\newcommand{\adross}{$\theta_{\rm Ross}$}
\newcommand{\adud}{$\theta_{\rm UD}$}
\newcommand{\adld}{$\theta_{\rm LD}$}

\bibliographystyle{apj}

\title{On the Limb Darkening, Spectral Energy Distribution, and Temperature Structure of Procyon}
\author{J. P.~ Aufdenberg\altaffilmark{1}}
\affil{National Optical Astronomy Observatory, 950 N. Cherry Ave, Tuscon, AZ 85719, USA}
\author{H.-G.~ Ludwig}
\affil{Lund Observatory, Lund University, Box 43, 22100 Lund, Sweden}
\and
\author{P.~ Kervella}
\affil{LESIA, UMR 8109, Observatoire de Paris-Meudon, 5 place Jules Janssen, 92195 Meudon Cedex, France}
\altaffiltext{1}{Michelson Postdoctoral Fellow}

\begin{abstract}
We have fit synthetic visibilities from 3-D ({\tt CO$^5$BOLD} + {\tt
PHOENIX}) and 1-D ({\tt PHOENIX}, {\tt ATLAS 12}) model stellar
atmospheres of Procyon (F5 IV) to high-precision interferometric data
from the VLTI Interferometer (K-band) and from the Mark III
interferometer (500 nm and 800 nm). These data sets provide a test of
theoretical wavelength dependent limb-darkening predictions.  The work
of Allende Prieto et al. has shown that the temperature structure from
a spatially and temporally averaged 3-D hydrodynamical model produces
significantly less limb darkening at 500 nm relative to the
temperature structure of a 1-D MARCS model atmosphere with a standard
mixing-length approximation for convection.  Our direct fits to the
interferometric data confirm this prediction. A 1-D {\tt ATLAS 12}
model with ``approximate overshooting'' provides the required
temperature gradient.  We show, however, that 1-D models cannot
reproduce the ultraviolet spectrophotometry below 160 nm with
effective temperatures in the range constrained by the measured
bolometric flux and angular diameter. We find that a good match to the
full spectral energy distribution can be obtained with a composite
model consisting of a weighted average of twelve 1-D model atmospheres
based on the surface intensity distribution of a 3-D granulation
simulation.  We emphasize that 1-D models with overshooting may
realistically represent the mean temperature structure of F-type stars
like Procyon, but the same models will predict redder colors than
observed because they lack the multicomponent temperature distribution
expected for the surfaces of these stars.
\end{abstract}

\section{Introduction}
The connection between the transport of energy in stellar atmospheres
and the limb darkening of stellar photospheres has been under
investigation for nearly 100 years in the case of the Sun.  
Since 1906 \cite[]{schwarzschild2}, the darkening of the solar limb has
been investigated to understand the roles of convection and radiation
in the transport of energy through the Sun's atmosphere.  Assuming
that the mass absorption coefficient of the solar atmosphere was both
wavelength and depth independent and that the angle-dependent
intensity could be replaced by the mean intensity at each depth,
Schwarzschild derived a center-to-limb intensity variation based on a
purely radiative equilibrium temperature structure
\begin{equation}
\frac{I(\tau=0,\cos\theta)}{I(0,1)} = 1 - \beta(1- \cos\theta) \qquad (\beta\ = 2/3),
\end{equation}
where $\tau$ is the optical depth, $\theta$ is the angle between the
line-of-sight and the emergent intensity, and $\beta$ is the linear
limb-darkening (LD) coefficient.  Schwarzschild showed this LD law to
be more consistent with contemporary observations than a LD law based
on an adiabatic equilibrium temperature structure.  \cite{milne1921}
improved on Schwarzschild's mean intensity approximation and found that
a radiative equilibrium temperature structure with improved flux
conservation yielded a LD coefficient of $\beta=3/5$ in better
agreement with observations.

The studies of Schwarzschild and Milne were completed before hydrogen
was recognized as the  principal constituent of the Sun's
atmosphere by \cite{payne1925}, before the work of \cite{a_unsold1930}
concerning the effects of hydrogen ionization on the stability of
radiative equilibrium against convection, and before the importance of
the bound-free and free-free opacity of the negative ion of hydrogen
was recognized by \cite{wildt1939}.  \cite{plaskett1936} inverted
solar LD observations to find a temperature structure inconsistent
with radiative equilibrium, suggesting convection has a significant
effect on limb darkening.  \cite{keenan1938} showed that this
conclusion was acceptable only if the adiabatic gradient began at
optical depths considerably less than $\bar{\tau} \sim 2$, contrary to
\citeauthor{a_unsold1930}'s calculations.  The reasoning was that
convection currents should transport little energy above the zone of
instability and that radiative equilibrium should prevail at smaller
optical depths significant to limb darkening.  Plaskett's conclusion
was further critiqued by \cite{woolley1941} who showed that it
necessitates convective velocities large enough to blur the Franuhofer
lines.  Woolley attributed the differences between the observations
and the radiative equilibrium models to the frequency dependent
opacity of the solar atmosphere, not yet included in the models. As
model atmospheres were improved with the incorporation of line
blanketing and mixing-length convection, subsequent studies
\cite[]{munch1945, swihart56, spiegel63, tm73, val76, kkk77} generally
confirmed \citeauthor{woolley1941}'s suggestion and concluded that the
effects of convection on limb darkening were subtle or
insignificant. Recently, \cite{cgk97} have shown that an ``approximate
overshooting'' modification to the standard one-dimensional (1-D)
mixing-length convection treatment enhances convective transport at
smaller optical depths and provides a better fit to solar LD
observations than models without overshooting.  
Overshooting refers
to the depth of convective penetration into layers of the atmosphere
stable against convection under the Schwarzschild criterion.

Photospheric convection and limb darkening are again the focus of some
of the most exciting observational and theoretical studies in
astrophysics.  Recently, space based photometric observations
\cite[]{most04} have put an upper limit on pressure mode oscillations
excited by turbulent convection in the atmosphere of the F-subgiant
Procyon.  Precise ground based photometric microlensing observations
are providing multi-band limb-darkening measurements of solar-like
stars \cite[]{abe03} and recent observations of this kind are
challenging model LD predictions for cool giants
\cite[]{fields03}. Increasingly precise long baseline interferometric
measurements at optical and near infrared wavelengths
\cite[]{wak04,onh04,perrin04} are providing new tests of LD
predictions.  Tests of these predictions are important because they
aid not only in the interpretation of interferometric data, but in the
studies of eclipsing binaries, transiting extrasolar planets
and their host stars.

Along with these recent observational developments, 3-D,
time dependent radiation hydrodynamical simulations \cite[]{sn98,r03}
have been produced to study solar granulation.  With the aid of such
models, the study of spectral line formation in solar granulation
\cite[]{agsk04} has recently significantly revised the solar elemental
oxygen abundance.  These simulations model in great detail the
transition from convection dominated transport to radiation dominated
transport near the surface of the solar atmosphere.  Self-consistent,
and not employing the usual 1-D mixing-length theory, these
simulations predict a spatially and temporally averaged temperature
profile for the solar atmosphere which samples upflows and downdrafts
across the model solar surface.  Similar models have also been
developed for cooler and more evolved stars
\cite[e.g.,][]{co5bold,lah02,robinson04}. For the Sun,
multi-wavelength LD observations are used to test the {\em
intensities} predicted by these models in addition to spectroscopic
tests of the {\em flux} predictions.  Qualitatively, multiwavelength
solar LD predictions from 3-D models \cite[fig. 3]{asplund99} and 1-D
models with overshooting \cite[fig. 4]{cgk97} are in reasonable
agreement with the observations of \cite{nl94}.  Both these studies
find that standard 1-D mixing-length theory predicts limb darkening
which is too strong, most noticeably in the blue (500 nm). 
A similar relationship between 3-D and 1-D limb-darkening predictions
has been established for Procyon \cite[][hereafter, AP02]{ap02}.
These predictions for Procyon have until now not been directly tested
by long baseline interferometric observations and this is the
principal motivation for this paper.

Procyon ($\alpha$ CMi, HR 2943, HD 61421), an F-type subgiant \cite[F5
IV-V,][]{gnw01} with a white dwarf companion, has been a prime target
for both observational and theoretical asteroseismological studies
\cite[]{procyon99,cdg99} because its visual orbital solution and
measured angular diameter provide well-constrained estimates of
its fundamental stellar parameters: effective temperature, gravity,
radius, and mass; most recently improved upon by \cite{girard2000} and
\cite{procyon_k04}.  The model of AP02 reproduces spectral line
shifts, signatures of convective motion, measured from very high
resolution (R$\simeq$200,000) spectroscopy of Procyon.  A comparison
between this 3-D model and a homogeneous, hydrostatic 1-D model
atmosphere from the MARCS code \cite[]{agke97, gben75}, with
mixing-length convection and the same effective temperature and
gravity, shows that at 1 \micron\ the normalized center-to-limb
predictions are quite comparable, while at 450 nm the 1-D intensity
profile is up to 20\% fainter than the 3-D intensity profile at
intermediate limb angles.  This LD contrast can be quantified with
respect to interferometric observations by comparing the predicted LD
corrections for the two models. The LD correction is a model dependent
scale factor between the angular size derived from the visibility data
assuming a uniformly bright stellar disk and the true wavelength
independent angular size corresponding to a physical radius \cite[see
e.g.,][]{dtb2000}.  From AP02, the monochromatic LD corrections at 450
nm are 1.081 (1-D model) and 1.064 (3-D model), a difference of 1.6\%.
In this paper we test this prediction by comparing both 1-D and 3-D
models directly with precise interferometric visibility measurements
obtained in two intermediate bands at 500 nm and 800 nm, and 
in the K-band at 2.2 \micron.

It is also interesting to compare the uncertainty in a typical
interferometric effective temperature estimate ($\simeq$100-200 K) with
the horizontal temperature variation expected from 3-D models.  The
AP02 3-D model for Procyon shows a R.M.S spatial temperature variation
of $\sim$500 K (8\%) at a depth defined by a Rosseland optical depth of unity. As
a result, while the bolometric flux and the angular diameter provide a
temperature which represents the bolometric output of the photosphere,
a 1-D atmosphere with this temperature is not expected to accurately
represent the spectral energy distribution.  For this reason, in
addition to testing the 3-D model limb-darkening predictions, we also
compare predictions of a 3-D model atmosphere to archival spectrophotometry
and Str\"{o}mgren photometry.

The broad wavelength coverage needed for the interferometric test
requires data from more than one interferometer.  We use 500 nm and
800 nm data from the Mark III interferometer \cite[]{shao88} and 2.2
\micron\ data from the Very Large Telescope Interferometer (VLTI),
specifically its commissioning instrument VINCI \cite[]{vinci04}.
In \S\ref{data} we describe the observations and present the
observational data and in \S\ref{models} we describe the model
atmospheres and the computation of the synthetic visibilities and the
synthetic spectrophotometry.  In \S\ref{results} we show the
comparison of the synthetic visibilities computed from the model
atmospheres with the visibility data and archival photometric
data. These results are discussed in \S\ref{discussion} and briefly
summarized in \S\ref{summary}.

\section{Observations and Fundamental Parameters}
\label{data}
\subsection{Interferometric Observations}
\subsubsection{VLTI Observations}
The European Southern Observatory's Very Large Telescope
Interferometer \cite[]{glindemann03} has been operated on top of the
Cerro Paranal, in northern Chile since March 2001. The new
observations presented in this paper were obtained by combining
coherently the light coming from the two VLTI test siderostats (0.35 m
aperture) on the E0-G1 baseline of the VLTI (66\,m in ground length).

We used the VINCI beam combiner \cite[]{kervella_vlti03} equipped with a
regular K band filter ($\lambda = 2.0-2.4\,\mu$m). While the
observations presented by \cite{procyon_k04} were limited to a
maximum baseline of 24\,m, the 22 new squared visibility measurements
cover the 42-66\,m range, significantly lower in the first lobe of the
Procyon visibility curve.  Both the 24\,m and 66\,m baseline data are
listed in Table~\ref{tab:vlti_data}.

The raw data were processed using a wavelet based algorithm
\cite[]{vinci04} to obtain the instrumental squared moduli of the
coherence factors.  The instrumental transfer function was deduced
from observations of four calibrator stars (HR\,1799, 6\,Leo, 18\,Mon
and 30\,Gem) whose {\it a priori} angular diameters were taken from
the \cite{cohen_1999} list. Their linear limb darkening coefficients
were taken from \cite{claret2000} tables for the K band. The 24\,m
baseline data previously published in \cite{procyon_k04}, as well as
three points from the 66\,m baseline, were calibrated using Sirius as
the reference star.  Its well-determined angular diameter
\cite[]{kervella_sirius_2003} results in a small systematic
uncertainty on the calibrated visibilities of Procyon.  Furthermore,
the Sirius-calibrated visibilities obtained on the longer E0-G1
baseline show no detectable deviation from the other calibrators.
The other calibrators have small angular diameters compared to Procyon
($\approx$2\,mas vs. more than 5\,mas). This translates into small
systematic errors on the calibrated visibility data. These errors were
carried through the data reduction process and are reported in
Table~\ref{tab:vlti_data}, together with the associated calibrator
stars.  

Apart from Sirius, the calibrators of the VLTI-VINCI data are all K
giants from the list established by \cite{cohen_1999}. For this
homogeneous set of stars, the limb darkening correction LD/UD is
always in the 1.02-1.03 range. The LD correction for Procyon in the K
band being just below LD/UD=1.02, it is almost identical and this
correction cancels out. For Sirius, the effective temperature is
higher ($\sim$10000 K), resulting in a smaller LD correction of
1.012. In order to take into account properly this 0.7\% difference,
we have used the linear LD correction from
\cite{claret2000}. Considering the small amplitude of this correction
(1.2\%), the introduced systematic uncertainty is considered
negligible compared to the VINCI error bar on the best estimate LD angular
diameter of Procyon, 5.404 $\pm$ 0.031 mas, $\pm$0.6\% (see \S\ref{results}).
The infrared K band is an advantage in this
matter, as the LD corrections are small, and do not change much over a
broad range of effective temperatures.

\subsubsection{Mark III Observations}
The Mark III interferometric data presented here comprise visibility
measurements at 800 nm \cite[fig. 1(i)]{m3_91} from which uniform disk
angular diameter fits have already been published \cite[]{m3_91,m3_03}
and unpublished visibility measurements at 500 nm from D. Mozurkewich
(2004, private communication).  We list these data in
Tables~\ref{tab:m3_500_data} and ~\ref{tab:m3_800_data} so that others
may fit models directly to these visibilities.  \cite{m3_91,m3_03} do
not provide a listing of calibrator stars. They note, however, that
the calibrator stars are all smaller than the program stars and we
expect the calibrator visibilities to be largely insensitive to
model-dependent systematic uncertainties due to limb darkening.
Seriously large systematic errors seem to be ruled about by a
comparison of the published Mark III uniform disk fits at 450 nm to
the earlier Intensity Interferometer uniform disk fits at 443 nm
\cite[]{hb74}, which agree within 1$\sigma$.

\subsection{Initial Fundamental Parameters}
\label{teff}
\subsubsection{Effective Temperature}
Fundamental stellar parameters for Procyon are well constrained by
direct measures of its bolometric flux and angular diameter.  The
bolometric flux is derived from an integration of Procyon's spectral
energy distribution (SED).  Absolute spectrophotometry in the
ultraviolet come from the {\em International Ultraviolet Explorer}
(low dispersion, large aperture spectra SWP43428 and LWR09108 from the
INES archive \cite[]{ines99}, effectively from 170 nm to 306 nm), and
from the {\em Hubble Space Telescope} Space Telescope Imaging
Spectrograph (10 one-dimensional extracted spectra in the sequence
o6l501020 to o6l5010k0 from 220 nm to 410 nm).  For integration, the
ultraviolet data are binned to 1 nm resolution.  Absolute
spectrophotometry in the visual and near-infrared wavelength range
(322.5 nm -- 1027.5 nm, 5nm resolution) has been obtained by
\cite{g92}.  For the SED beyond 1 micron, the broad-band photometry
(JHKLMN) from \cite{mm78}, with the the absolute calibration of
\cite{j65} for JKLMN and \cite{bb88} for H, is fit using linear
interpolation in $\log \lambda - \log F_\lambda$ space.  We
have adopted a uniform flux uncertainty of 5\% for all the
spectrophotometric data.  The most carefully calibrated photometric
standards are not likely to be more accurate than 4\% in the absolute
sense \cite[]{bg_vega04}.  On the basis of Procyon's distance (3.5
pc), interstellar dust extinction is assumed to be negligible.  The
integrated flux from 170 nm to 10200 nm is $17.82\pm0.89\times
10^{-9}$ W m$^{-2}$.  This value is in good agreement with the value
$18.08\pm 0.76$ W m$^{-2}$ derived by \cite{code76}.  The difference
in the integrated flux between STIS and {\em IUE} over their common
wavelength range is less than 1\%.  The integrated STIS flux is $\sim
6\%$ larger than the integrated \cite{g92} flux between 322 nm and 410
nm.  Our integrated flux and the limb-darkened angular diameter
(\adld\ = 5.45$\pm$0.05 mas) from \cite{procyon_k04} yield \teff\ =
$6516 \pm 87$ K.  This value is in good agreement with \teff\ =
$6510\pm 150 $ K from \cite{code76}.  A comparison of the angular
diameter uncertainty (0.9\%) and the integrated flux uncertainty (5\%)
shows that the uncertainty in \teff\ is dominated by the
uncertainty in the spectrophotometry.

\subsubsection{Radius and Surface Gravity}
The {\it HIPPARCOS} parallax \cite[285.93$\pm$0.88 mas,][]{hipparcos}
and the angular diameter (\adld\ = 5.45$\pm$0.05 mas) yield a radius
of 2.05$\pm$0.02 \rsun.  The orbital solution of \cite{girard2000}
provides a mass for Procyon A, which is 1.42$\pm$0.04 \msun\ adopting
the {\it HIPPARCOS} parallax \cite[]{procyon_k04}.  The mass together
with the radius yields a surface gravity, $\log(g)$ =
3.95$\pm$0.02 (in cgs units), which is uncertain at the 5\% level.

\section{Model Atmospheres}
\label{models}
The construction of model atmospheres for Procyon and the subsequent
calculation of synthetic radiation fields for comparison with
interferometry and photometry is done three ways: (1) Stand-alone 1-D
{\tt PHOENIX} \cite[]{nextgen2} structures, radiation fields and
spectra ; (2) {\tt CO$^5$BOLD} \cite[]{co5bold} 3-D structures
temporally and spatially averaged to 1-D, then read by {\tt PHOENIX}
for computation of the corresponding radiation fields; (3) Stand-alone
1-D {\tt ATLAS 12} \cite[]{atlas12} structures, radiations fields and
spectra.  These models will be discussed in more detail below.

\subsection{Input Parameters}
The {\tt ATLAS 12} models are plane parallel and require an effective
temperature \teff\ and a surface gravity, $\log(g)$, as input
parameters.  The spherical {\tt PHOENIX} models additionally require a
stellar radius because in the spherical case the luminosity, not the
flux, is constant. For Procyon's photosphere this distinction is
negligible, however the boundary radii of the spherical and
plane-parallel models differ as discussed below.  {\tt CO$^5$BOLD}
models a local, 3-D, statistically representative volume of the
stellar atmosphere in Cartesian geometry with periodic lateral
boundaries.  As input parameters {\tt CO$^5$BOLD} requires the surface
gravity and the specific entropy of material entering the
computational volume through the open lower boundary. The entropy
strongly influences the amount of heat entering the volume from below,
and plays a role analogous to the effective temperature in the 1-D
models.

\subsection{{\tt PHOENIX} Models}
A grid of forty-two 1-D, spherical, hydrostatic models from the {\tt
PHOENIX} general-purpose stellar and planetary atmosphere code
\cite[version 13.07, for a description see][]{nextgen2,hb99} have been
constructed with the following parameters: effective temperatures,
\teff\ = 6420 K to 6620 K, 10 K steps; $\log(g) = 3.95$, radius, $R =
1.46\times 10^{11}$ cm; depth-independent micro-turbulence, $\xi$ =
2.0 \kms (a value suggested for 1-D models by AP02 from their analysis
of Procyon's spectrum); mixing-length to pressure scale ratio,
$\alpha$ = 0.5 and 1.25; solar elemental abundance \cite[]{solab89};
LTE atomic and molecular line blanketing, typically $10^6$ atomic
lines and $4\times 10^{5}$ molecular lines dynamically selected at run
time; non-LTE line blanketing of \ion{H}{1} (30 levels, 435
bound-bound transitions), \ion{He}{1} (19 levels, 171 bound-bound
transitions), and \ion{He}{2} (10 levels, 45 bound-bound transitions);
and boundary conditions: outer pressure, $P_{\rm gas} = 10^{-4}$ dynes
cm$^{-2}$, extinction optical depth at 1.2 \micron: outer $10^{-10}$,
inner $10^{2}$.  For this grid of models the atmospheric structure is
computed at 100 radial shells (depths) and the radiative transfer is
computed along 100 rays tangent to these shells and 14 additional rays
which impact the innermost shell, the so-called core-intersecting
rays. The intersection of the rays with the outermost radial shell
describes a center-to-limb intensity profile with 114 angular points.
The mixing-length theory for convection in {\tt PHOENIX} is very
similar to the Mihalas formulation \cite[]{lah02,mihalas78} with no
overshooting.  A second grid of 24 {\tt PHOENIX} models has been
constructed as components to composite models which simulate the
effects of granulation on the spectral energy distribution (see
\S\ref{synthetic_colors}, Table~\ref{tab:component_colors}).  Other
than \teff, these models have the same input parameters as the first
grid.  The non-LTE H and He aspect the stand-alone {\tt PHOENIX}
models is an insignificant factor for the angular diameter fits and
synthetic spectrophometry presented here, as we have confirmed by
converging purely LTE {\tt PHOENIX} models for comparison.  The H$^-$
ion is treated in LTE for all the {\tt PHOENIX} models.  The effects
of treating many thousands of lines out of LTE in a solar-type star
has been recently explored by \cite{sh05}.  Such effects could be
important for the synthetic spectrophotometry, however such
massive-scale non-LTE models for Procyon have yet to be calculated and
are beyond the scope of this paper.  Below we specifically refer to
{\tt PHOENIX} models ``A'' and ``B''.  For both models \teff = 6530 K
and $\log(g) = 3.95$ (consistent with the values derived in
\S\ref{teff}), but they have different values for $\alpha$, 1.25 and
0.5, respectively.

\subsection{{\tt CO$^5$BOLD} and {\tt PHOENIX} Models}
The ``COnservative COde for COmputation of COmpressible COnvection in
a BOx of L Dimensions ($L=2,3$)'' ({\tt CO$^5$BOLD}) was mainly
developed by B.~Freytag and M.~Steffen \cite[for details
see][]{co5bold,wedemeyer04}.  The {\tt CO$^5$BOLD} models (series
codes d3gt65g40n2.80-a4-3 (grey) and d3gt65g40n3.10-26-3 (non-grey))
for Procyon have an average \teff\ = 6500~K, a prescribed $\log(g)$ =
4.0, and solar abundances.  For the non-grey model, five wavelength
groups are employed to describe the wavelength dependence of the
radiation field within a multi-group radiative transfer scheme
\cite[]{ludwig_thesis}. The group-averaged opacities are based on data
provided by the {\tt ATLAS~6} code \cite[]{atlas6}.  

Integration with the 1-D {\tt PHOENIX} code provides direct opacity
sampling at high spectral resolution for calculation of the radiation
field based temperature structures provided by {\tt CO$^5$BOLD}.  The
1-D temperature, pressure, and depth arrays (91 depth points) from
{\tt CO$^5$BOLD} are read by {\tt PHOENIX}.  The physical depths
provided by {\tt CO$^5$BOLD} are relative to the layer where the
average Rosseland optical depth is unity, but {\tt PHOENIX} requires
absolute radii for construction of the 1-D spherical radiation field.
The value $1.46\times 10^{11}$ cm, corresponding to the physical
radius of Procyon, is therefore added to the {\tt CO$^5$BOLD} depths.
The accuracy of this value is not critical because the thickness of
the atmosphere is $\sim 0.1$\% of the stellar radius.

\begin{figure}
\includegraphics[scale=0.8,angle=0]{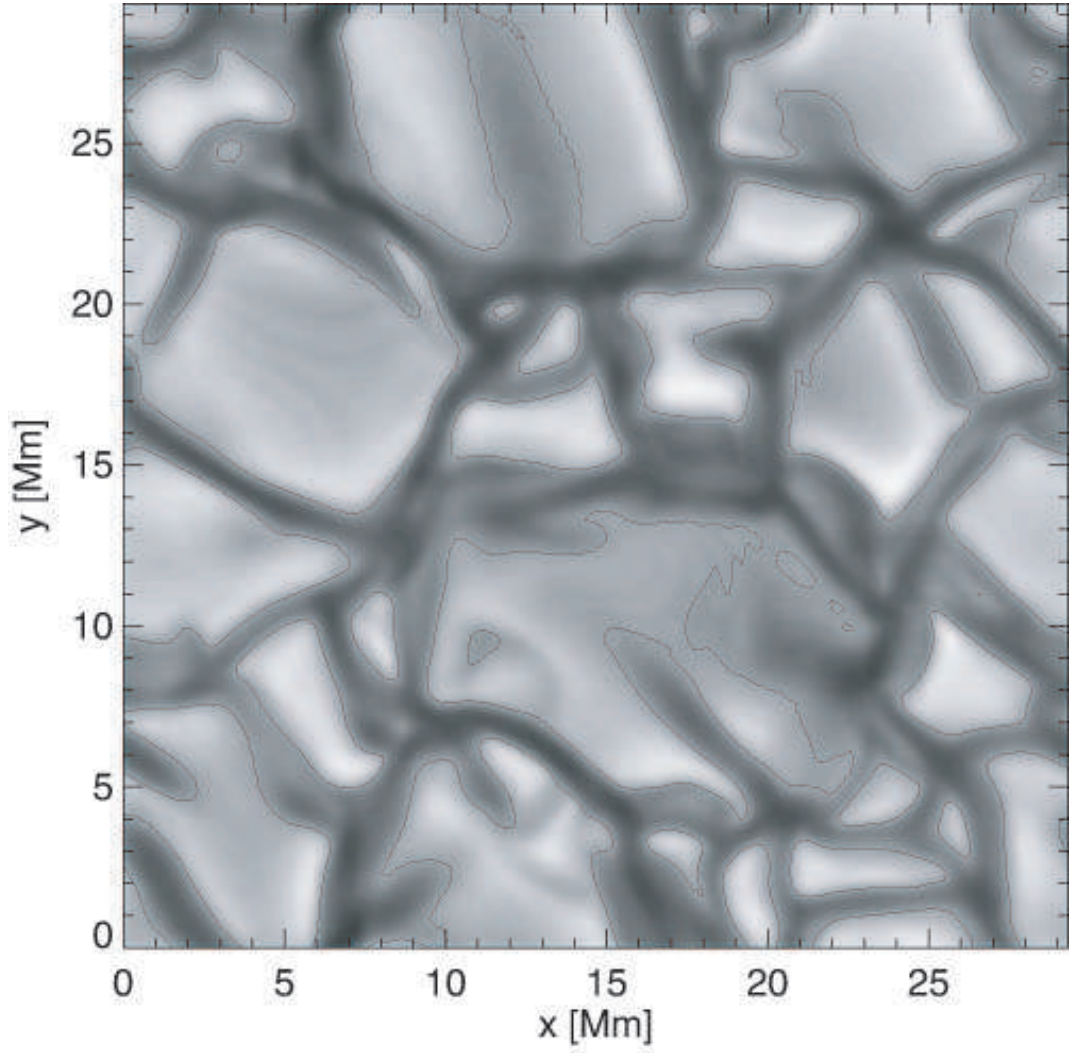} 
\caption{ Typical map of the emergent white light intensity in
   vertical direction
   encountered during the temporal evolution of the grey {\tt
   CO$^5$BOLD} model. The black line depicts an isophote at
   $3.95 \times 10^{10}$\,\mbox{erg\,cm$^{-2}$\,sr$^{-1}$\,s$^{-1}$}
   approximately corresponding to the image's median intensity (see
   Figure~\ref{fig:intensity_histogram}).}
\label{fig:intensity_map}
\end{figure}

\begin{figure}
\includegraphics[scale=0.8,angle=0]{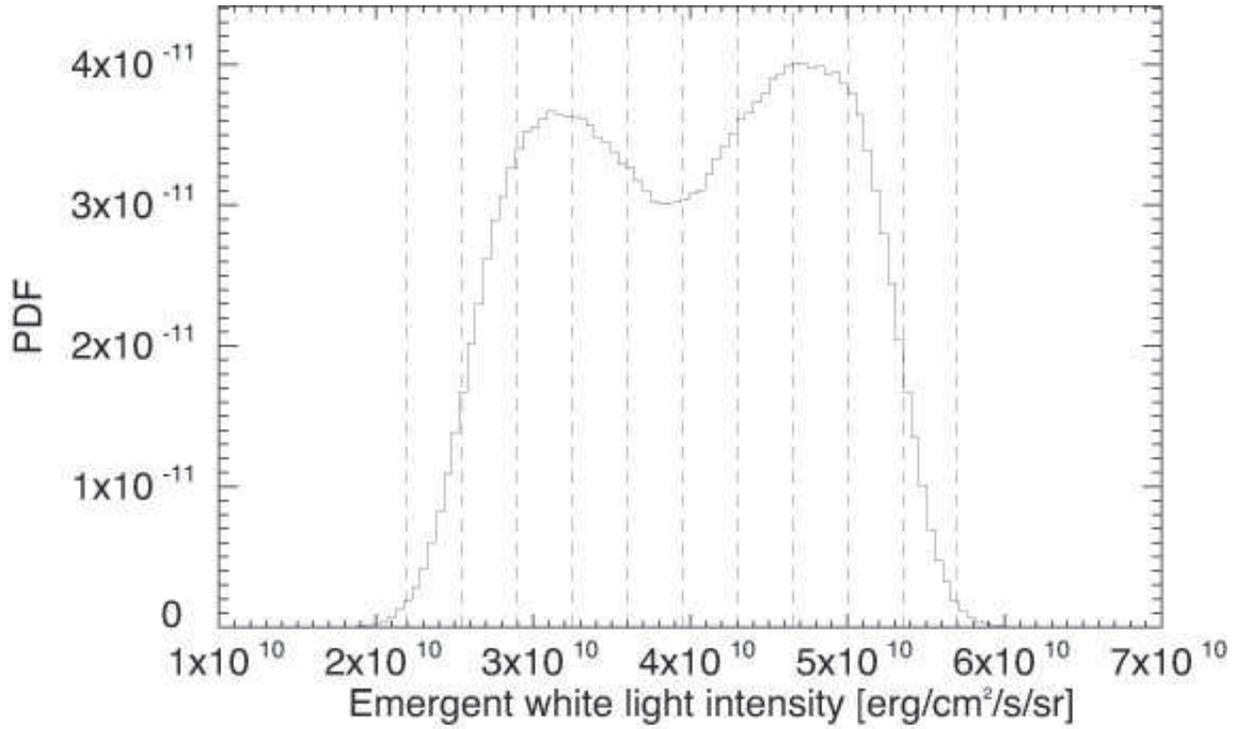} 
\caption{Probability density function (PDF), the probability per
differential intensity interval with the units of inverse intensity,
of the emergent white light intensity of the grey {\tt CO$^5$BOLD}
model. The PDF was derived from a sequence of intensity maps like the
one shown in Figure~\ref{fig:intensity_map}.  Twelve intensity
intervals are demarcated by the dashed lines.  The surface area
weights for these intervals are given in
Table~\ref{tab:component_colors}.  }
\label{fig:intensity_histogram}
\end{figure}

We take two approaches for the computation of the mean radiation field
of the {\tt CO$^5$BOLD} models. (1) we compute a {\tt PHOENIX}
radiation field from a temporally and spatially averaged 3-D
temperature structure. (2) we compute 980,000 horizontal positions
from the spatial and temporal evolution of the granular flow (see
Figure \ref{fig:intensity_map} for a snapshot of the flow) and sort
them into 12 groups according to their intensity (see Figure
\ref{fig:intensity_histogram} for the histogram of intensities). Next,
we average the vertical slabs on surfaces of fixed optical depth.
This results in twelve temperature stratifications representing the
dark and increasingly brighter granulation areas.  {\tt PHOENIX}
radiation fields are then computed for each of the twelve structures.
These radiation fields are then co-added, weighted by the surface
areas of the twelve intensity groups (for weights see
Table~\ref{tab:component_colors}).  Theoretically, we expect approach
(2) provides a better match to the actual radiation field of Procyon.
This is because radiative transfer in an inhomogeneous medium is
non-linear and one cannot interchange radiative transfer and spatial
averaging exactly. In this respect, the twelve component model should
be an improvement over a global spatial mean.

We find the temporally and spatially averaged temperature structures
from the grey and non-grey {\tt CO$^5$BOLD} models are nearly
identical at the depths of continuum formation for the interferometric
wavebands considered here: 500 nm, 800 nm, and 2.2 \micron.  As a
result, the corresponding {\tt PHOENIX} radiation fields from approach
(1) yield fit angular diameters that differ by less than $0.03$\%.
Much greater than the grey versus non-grey effects are the differences
using approaches (1) and (2).  Using the weighted average of the
twelve radiation fields to simulate the mean center-to-limb profile
yields an angular diameter 0.4\% larger at 500 nm compared to the
global spatial mean radiation field.  As we show in \S\ref{results},
approach (2) is slightly more consistent with the observed visibility
data.  Below we refer to models ``C1'' and ``C2'', models constructed
via approachs (1) and (2), respectively.

Tests indicate that the {\tt PHOENIX} radiation field produced from
the {\tt CO$^5$BOLD} 1-D average structure produces an outer boundary
radiative flux lower than expected (by 15\% in flux, 260 K in \teff)
for an effective temperature of 6500 K.  This is most likely due to
the different opacity setups in the two codes.  {\tt CO$^5$BOLD} uses the
older {\tt ATLAS 6} opacity setup.  As a result, a spectral energy
distribution from {\tt PHOENIX} based on the mean 3-D structure will
be cooler than expected from the {\tt CO$^5$BOLD} \teff\ value.  This is not
a critical problem for the interferometric comparisons because of the
insensistivity of the limb-darkening to \teff\ in this range (5500 K
to 7500 K, see \S\ref{gradient}).  The radiative flux mismatch is a
problem with regards to synthetic photometry and this is discussed
below in \S\ref{synthetic_colors}.

\subsection{{\tt ATLAS 12} Models}
We also use {\tt ATLAS 12} models \cite[and R.~L. Kurucz 2004, private
communication]{atlas12} which include an ``approximate overshooting''
(AO) prescription for convective flux transport in a mixing-length
formalism \cite[sec. 2.3]{cgk97}, for comparison with the 1-D and 3-D models
just discussed.  In the AO formulation, the convective flux extends to
lower optical depths, reducing the temperature gradient relative to
models without overshooting.  Three {\tt ATLAS 12} models have been
constructed: ``D'', no overshooting; ``F'', 50\% overshooting; and
``E'' full overshooting.  All three models have the following
parameters: \teff\ = 6530 K; $\log(g) = 3.95$; the microturbulence is
depth dependent increasing from 0.9 \kms\ to 3.2 \kms\ at the inner
boundary; $\alpha$ = 1.25; solar elemental abundance \cite[]{solab89};
LTE atomic and molecular line blanketing; and boundary conditions:
outer $\tau_{\rm Ross} = 10^{-7}$, inner $\tau_{\rm Ross} = 10^{2}$.
After the models, with 72 depth points, were computed by opacity
sampling, the complete spectrum was computed at resolving power of
R=500,000 for 17 angular ``$\mu$'' points describing the
center-to-limb intensity profile.

\subsection{Synthetic Visibilities and Fit Procedure}
The computation of the synthetic visibilities from the model radiation
fields simulates the bandwidth-smeared squared visibility 
\cite[see e.g.,][]{dtb2000}.  At a projected baseline $B$ and mean wavenumber
$\lambda_0^{-1}$, the synthetic squared visibility,
\begin{equation}
V(B,\lambda_0)^2 = \frac{\int_0^{\infty} V(B,\lambda)^2\, \lambda^2 \, d\lambda}
{\int_0^{\infty} S(\lambda)^2\ F_{\lambda}^2\ \, \lambda^2 \, d\lambda},
\end{equation}
is computed at each wavelength from a Hankel transform,
\begin{equation}
V(B,\lambda) = \int_0^1 S(\lambda) I(\mu,\lambda)
J_0\biggl[\pi\theta_{\rm LD}(B/\lambda)(1-\mu^2)^{1/2}\biggr]\ \mu\, d\mu ,
\end{equation}
where $I(\mu,\lambda)$ is the model radiation field (in photons
cm$^{-2}$ s$^{-1}$ sr$^{-1}$), $S(\lambda)$ is the instrument sensitivity
curve, $\mu$ is the cosine of the angle between the line of sight and
the surface normal, and $\theta_{\rm LD}$ is the limb-darkened angular
size.  The mean wavenumber is computed from
\begin{equation}
\lambda_0^{-1} = \frac{\int_0^{\infty}  \lambda^{-1}\ S(\lambda)\ F_{\lambda}\ \, d\lambda}
{\int_0^{\infty}  S(\lambda)\ F_{\lambda}\  \, d\lambda} ,
\end{equation}
where 
\begin{equation}
F_\lambda = 2\pi\ \int_0^1 I(\mu,\lambda)\ \mu\ \, d\mu
\end{equation}
is the flux.  For each model radiation field, non-linear least squares fits are
performed to the visibility data at 500 nm, 800 nm, and 2.2 \micron,
using the Interactive Data Language (IDL) routine {\tt CURVEFIT},
yielding three values for $\theta_{\rm LD}$.

The baselines for the VINCI visibilities are listed in
Table~\ref{tab:vlti_data}.  The reciprocal of the mean wavenumber for
the VINCI transmission (including: K-band filter, detector quantum
efficiency, fibers, atmosphere, and MONA beam combiner) is
$\lambda_0$ = 2.182$\pm$0.002 \micron\ using an appropriate synthetic
spectrum for Procyon (a \teff=6530 K, $\log(g)=3.95$ {\tt PHOENIX}
model). An accurate estimate for the mean wavenumber is important
because the fit angular diameter scales linearly with the mean
wavenumber.

The Mark III sensitivity curves are assumed to be Gaussian with
central wavelengths $\lambda_0$ = 500 nm and 800 nm, each with a FWHM
of 20 nm \cite[]{m3_91}.  The Mark III bands are $\sim$20 times
narrower than the VINCI band, therefore the mean wavenumber is much
less sensitive to the shape of Procyon's spectral energy distribution.

\subsubsection{Accounting for Extension in the Spherical Models} 
A proper comparison of the best fit $\theta_{\rm LD}$ values derived
from {\tt PHOENIX}, {\tt CO$^5$BOLD} +{\tt PHOENIX}, and {\tt ATLAS
12} models requires a correction to the stand-alone {\tt PHOENIX}
values because of spherical extension effects.  The {\tt PHOENIX}
structures extend outward to a radial shell $\simeq 0.4$\% above the
$\tau_{\rm ross} = 1$ radius while the model plane-parallel structures
have very small extensions for a fixed gravity (the {\tt CO$^5$BOLD}
and {\tt ATLAS 12} structures have extensions of $\simeq 0.1$\%).  The
radiation field $I(\mu,\lambda)$ refers to the angular distribution of
intensities emerging from the outermost model layer.  As a result, the
spherical models with greater radial extension yield larger angular
sizes relative to the less extended plane-parallel models.  This is
true even when the angular size of the continuum-forming radii for the
two model geometries agree because their outer boundary radii differ.
Thus, the {\tt PHOENIX} $\theta_{\rm LD}$ values have been scaled down
to correspond with the $\tau_{\rm ross} = 1$ radius (in other words
reduced by 0.4\%) for comparison with the plane-parallel models.  For
red giants the correction is much larger, for example 7\% in the
case of $\psi$ Phe \cite[]{wak04}.  This correction reminds us that
even relatively compact atmospheres are ``fuzzy'' and that the stellar
radius must be carefully defined \cite[]{baschek91}.  The need for
this correction also speaks to the precision of the observational
data.

\subsection{Synthetic Spectrophotometry and Photometry with 3-D and 1-D Models}
\label{synthetic_colors}
At present, the coarseness of the wavelength-dependent opacity in the
{\tt CO$^5$BOLD} model does not permit direct predictions of accurate
colors.  Therefore, we use the spatial intensity distribution of the
3-D model as a guide to construct multicomponent models of Procyon's
granulation pattern with {\tt PHOENIX}.  From the grey {\tt
CO$^5$BOLD} model a snapshot of the emergent intensity in integrated
(``white'') light is shown in Figure~\ref{fig:intensity_map},
Figure~\ref{fig:intensity_histogram} depicts a histogram of
intensities, including 12 intensity intervals for which we construct
{\tt PHOENIX} models (see Table~\ref{tab:component_colors} for
corresponding surface area weights). The bimodal nature of the
histogram shows that the model photosphere can be divided roughly into
``cool'' and ``hot'' components, as also indicated in the intensity
map. In fact, we initially experimented with such a two-component
model, and later refined the spectral synthesis to the present
12-component model.

For constructing the {\tt PHOENIX} models we need to specify a
\teff. Assuming a linear darkening law
\begin{equation}
\frac{I_\lambda(\mu)}{I_\lambda(1)} = 1 - \beta(1-\mu)
\end{equation}
we convert emergent intensities in vertical direction
to fluxes~$F$ according to
\begin{equation}
F_\lambda = \pi I_\lambda(1) \left(1 - \frac{\beta}{3}\right).
\end{equation}
With the additional assumption of a wavelength independent $\beta$ we
directly obtain \teff\ since the grey {\tt CO$^5$BOLD} model provides
wavelength integrated intensities.  The \teff\ of the inner 10 zones
are taken as the average of the \teff\ at the two adjacent boundaries.
For the two outer zones half the interval width $\Delta\teff$ of the
adjacent zone is added (subtracted) to the hottest (coolest)
boundary. The radiation fields of the individual components are added,
weighted by the surface area fraction associated with the represented
intensity interval. The resulting \teff\ values and weights for the 12
zones are listed in Table~\ref{tab:component_colors}.  Two values for
$\beta$ have been chosen: 0.5 and 0.6.  These values are consistent
with published linear limb-darkening coefficients for broad-band
optical filters \cite[]{claret2000}. More importantly, these two
$\beta$ values yield \teff\ distributions which, after the twelve 1-D
model SEDs are weighted and co-added, have bolometric fluxes and
corresponding effective temperatures which bracket the {\tt
CO$^5$BOLD} model (\teff\ = 6500 K). The composite spectrum \teff\ is
computed as
\begin{equation}
T_{\rm eff} = \left\{ \frac{1}{\sigma} \int_0^{\infty} \left[\sum_{i=1}^{i=12} w^i F^i_\lambda \right] \, d\lambda \right\}^{1/4}
\end{equation}
where $\sigma$ is the Stefan-Boltzmann constant, $w^i$ are the weights
given in Table~\ref{tab:component_colors}, and $F^i_\lambda$ are the
flux distributions from the component models.

\subsubsection{Synthetic Photometry}
\label{zero-points}
Synthetic Str\"{o}mgren indices \cite[]{stroemgren66} are computed
using the formulae of \cite{matsushima69}.  In this formalism, a
normalization constant, or zero point, $k_{ij}$ is needed to transform each
natural index, 
\begin{equation}
(i-j)_{\rm nat} = 2.5 \log \left\{\frac{\displaystyle \int_0^{\infty} F_\lambda S_j(\lambda)\, d\lambda}
{\displaystyle \int_0^{\infty} F_\lambda S_i(\lambda)\, d\lambda} \right\}
\end{equation}
to a photometric index,
\begin{equation}
(i-j) = (i-j)_{\rm nat} + k_{ij}
\end{equation}
directly comparable to
observations.  In the literature, the normalization of the theoretical
indices is often based solely on models of Vega \cite[e.g.,][]{sk97},
however \cite{lgk86} have shown that these zero points are a function
of $(b-y)$.  Since Vega and Procyon do not have the same $(b-y)$, using
Vega leads to systematic errors in the synthetic indices.  We also wish
for obvious reasons not to use a model of Procyon to obtain these
zero points.  Therefore, we use two independent spectrophotometric
observations of Procyon \cite[]{kiehling87,g92} and two different sets
of filter sensitivity curves \cite[]{matsushima69,cb70} to determine
the following zero points:
\begin{eqnarray}
{(b-y)}_{\rm obs}   - {(b-y)}_{\rm nat}    &= &+0.503 \pm 0.004 \pm 0.007 \\
{(c_1)}_{\rm obs}   - {(c_1)}_{\rm nat}    &= &-0.088 \pm 0.003 \pm 0.009 
\end{eqnarray}
These zero points are given with two sets of uncertainties.  The first
set arises from the range of zero points obtained using the four
different filter/spectrophotometric data combinations.  The second set
is from to the mean observational error in the observed indices
\cite[]{cb70}. If we use spectrophotometry of Vega
\cite[]{bohlin_vega96} the mean $c_1$ zero point is instead --0.165
mag, consistent with the $c_1$ shift shown by \cite{lgk86} at the
$(b-y)$ color (0.272) of Procyon.

\section{Results}
\label{results}
\subsection{Multiwavelength Angular Diameter Fits}
\begin{figure}
\includegraphics[scale=0.7,angle=0]{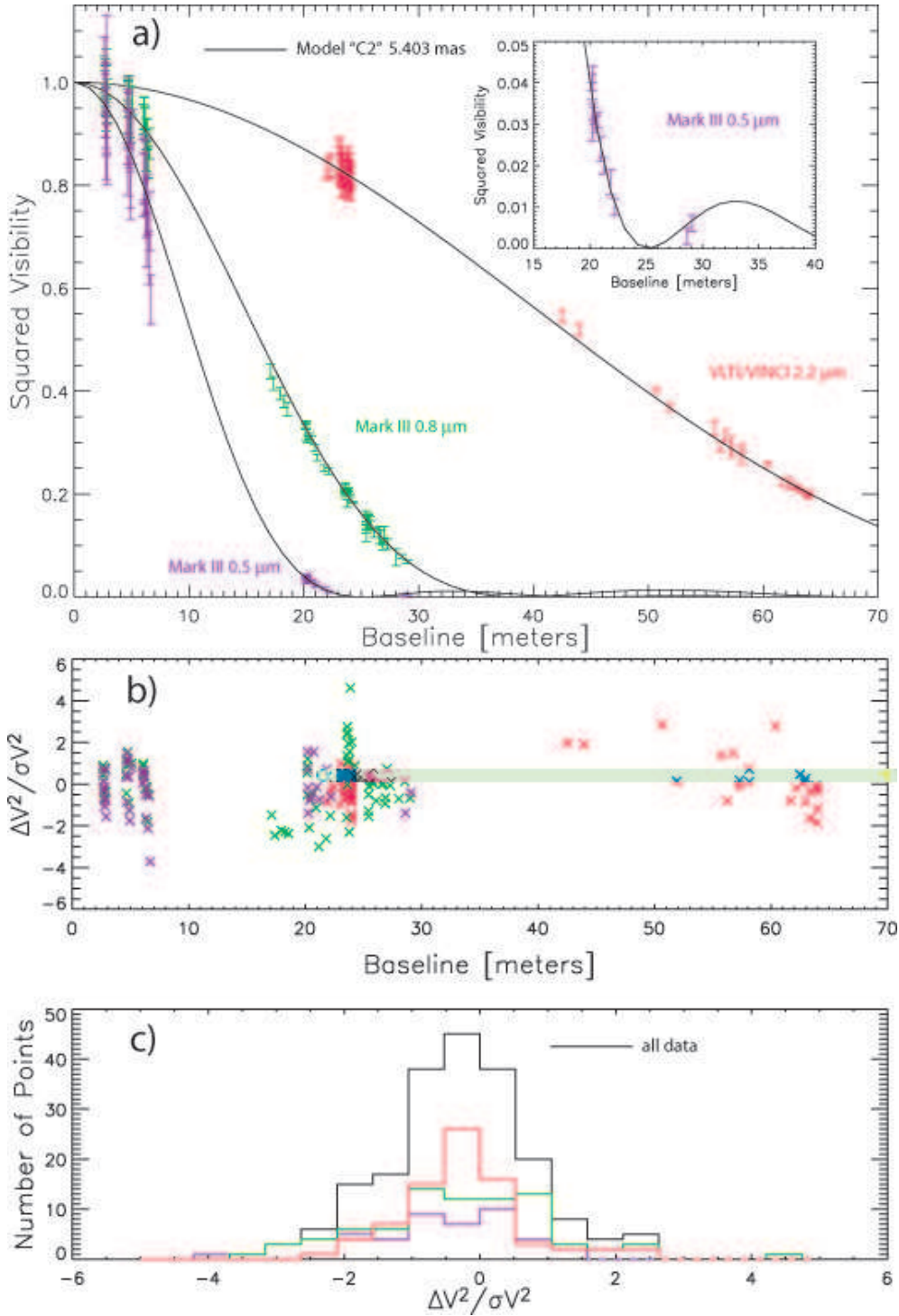}
\caption{(a) The squared visibility data from Tables
\ref{tab:vlti_data} (red), \ref{tab:m3_500_data} (purple) and
\ref{tab:m3_800_data} (green) as a function of projected baseline compared to
the synthetic visibilities (from eq.(2)) from the 3-D {\tt CO$^5$BOLD
+ PHOENIX} model ``C2'' with an angular diameter of 5.403
mas.  (b) The deviations of each data set from the
model. (c) Histograms of these deviations for each data set and for
all the data combined.}
\label{fig:v2_plot}
\end{figure}

\begin{figure}
\includegraphics[scale=0.6]{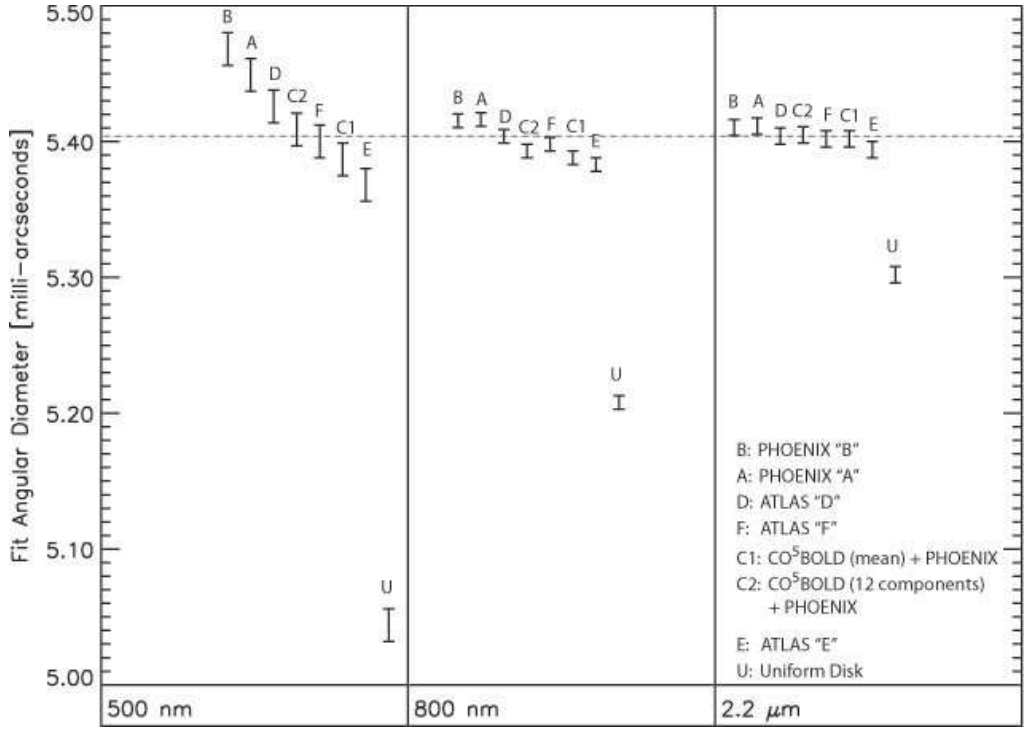}
\caption{A comparison of the best fit angular diameters $\theta_{\rm
LD}$ at 500 nm, 800 nm (Mark III data), and 2.2 \micron\ (VLTI data)
from Table~\ref{tab:AD_fits} for the seven atmosphere models ``A'' to
``F'' and one uniform disk model ``U''.  The uniform disk angular
diameters indicate that Procyon is more limb darkened at shorter
wavelengths as expected.  The atmosphere model fits are more dispersed
at shorter wavelengths.  The fits indicate that 1-D models without
overshooting are too limb darkened at 500 nm, while 1-D model ``E''
with 100\% overshooting is not limb darkened enough.  Model ``F'' with
50\% overshooting yield the same angular diameter (within 1$\sigma$)
at all three wavelengths. The 3-D {\tt CO$^5$BOLD} + {\tt PHOENIX}
models ``C1'' and ``C2'' yield angular diameters slightly less consistent than
model ``F'', but with no free parameters for convection.  The dashed
line indicates the angular diameter of the model in 
Figure~\ref{fig:v2_plot}.}
\label{fig:ad_diffs}
\end{figure}

Table~\ref{tab:AD_fits} lists the best fit $\theta_{\rm LD}$ values 
(and 1$\sigma$ uncertainties) at
500 nm, 800 nm, and 2.2 \micron\ for seven different atmospheric
structures and a uniform disk model.  Figure \ref{fig:v2_plot}
shows all the visibility data together with the synthetic visibilities
from the {\tt CO$^5$BOLD} + {\tt PHOENIX} model ``C2'' with the mean
fit angular diameter of 5.403 mas.  The best fit angular diameters for
all seven models are compared in Figure \ref{fig:ad_diffs}.  

There is a clear trend in the fit results toward shorter wavelengths.
The seven models yield a significantly wide range of angular diameters
at 500 nm, differing by 100$\pm$23 $\mu$as, while differing only by
17$\pm$12 $\mu$as at 2.2 \micron.  Of the seven models, two models stand
out as best representing the visibility data at all three wavelengths
with a single angular diameter: {\tt CO$^5$BOLD} + {\tt PHOENIX} model
``C2'' and {\tt ATLAS 12} model ``F'' (with 50\% overshooting). Both
models yield angular diameters at the 500 nm and 2.2 \micron\ that are
the same within the formal uncertainties: 15$\pm$17 $\mu$as and
4$\pm$17 $\mu$as, respectively.  Unlike model ``F'', model ``C2'' has
no free parameters for convection.

\subsection{Model Comparisons to Spectrophotometry and Photometry}
\begin{figure}
\includegraphics[scale=0.7,angle=0]{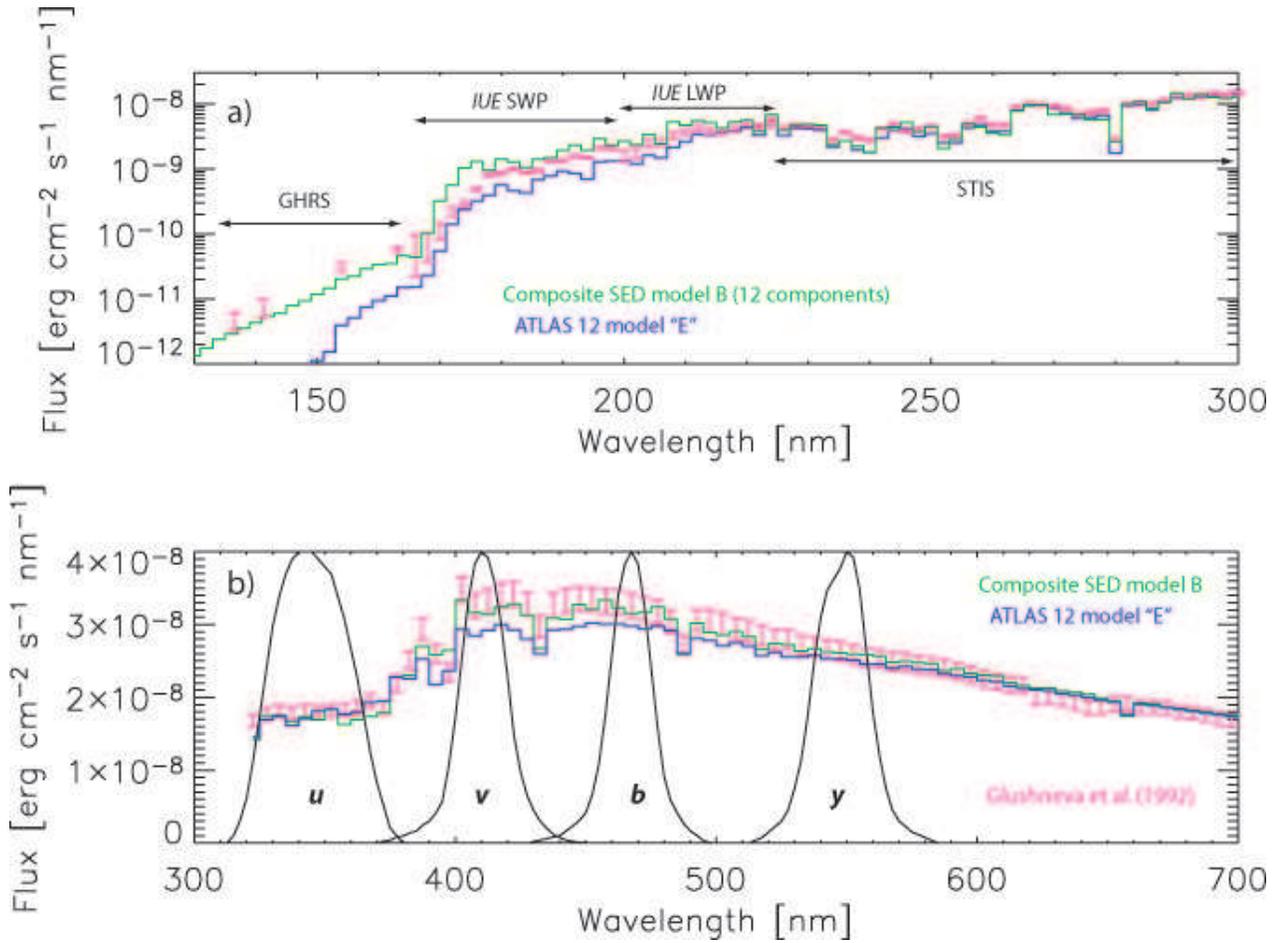}
\caption{Comparisons between synthetic spectral energy distributions
and spectrophotometric measurements at ultraviolet and visual
wavelengths. The models and data are binned to 2 nm resolution in the
UV for clarity. The models are binned to 5 nm in the visual to match
the resolution of the spectrophotometry.  The synthetic SEDs are
scaled in absolute flux using an angular diameter of 5.40 mas, a value
consistent with our best interferometric estimate (see
Table~\ref{tab:fund_params}).  (a) Composite SED model B (green),
which incorporates a 3-D model for the distribution of surface
intensities due to granulation, better represents the UV flux
distribution than a single \teff\ component 1-D model (blue, {\tt
ATLAS 12} model ``E''), particularly at wavelengths below 160 nm,
where all five single \teff\ component, 1-D atmosphere models fail.
(b) Composite SED model B (green) yields Str\"{o}mgren $v$- and $b$-band
fluxes consistent with the observed spectrophotometry \cite[]{g92} and
larger than the 1-\teff\ component model (blue, {\tt ATLAS 12} model
``E''). These two models are in closer agreement in the $u$- and
$y$-bands. }
\label{fig:sed_comparison}
\end{figure}

Figure \ref{fig:sed_comparison}(a) shows that the synthetic SED of
{\tt ATLAS 12} model ``E'' fails to reproduce the observed continuum
below 160 nm.  This is true not only for model ``E'', but all 1-D
models, regardless of convection treatment, which are consistent with
both the measured angular diameter and the measured bolometric flux.
While the contribution to Procyon's bolometric flux from radiation
between 130 nm and 160 nm is negligible, the absolute flux level in
this region provides strong evidence for a multicomponent temperature
surface distribution and therefore granulation.

The absolute continuum fluxes between 136 nm and 160 nm are from the
Goddard High Resolution Spectrograph data sets: Z2VS0105P (PI
A. Boesgaard), Z17X020CT, Z17X020AT, Z17X0208T (PI J. Linsky).  These
spectra were originally obtained to study the Procyon's atmospheric
boron abundance \cite[]{cunha2000} and chromosphere \cite[]{wood96}.
The signal-to-noise in the continua of these spectra ranges from
approximately 4:1 in the bluest data set to 10:1 in the reddest data
set.  These data are far superior to any other measurements below 160
nm because the continuum drops by more than a factor of 100 here, too
much for the limited dynamic range of {\it IUE}.  The absolute fluxes
below 160 nm shown in Figure \ref{fig:sed_comparison} were estimated
by computing the mean flux between the emission lines in each spectrum
incorporating the flux uncertainties provided with each calibrated
data set.

The observed continuum below 160 nm is consistent with a spatial
component of Procyon's atmosphere which has a color temperature in
excess of the interferometric \teff.  Therefore, such a component must
represent a small fraction of the surface, for it would otherwise
produce a bolometric flux in excess of that observed.  The continuum
of Procyon between 136 nm and 160 nm should be photospheric, in
contrast to the solar spectrum where the bound-free opacity of
silicon causes the continuum to form at or slightly above the
temperature minimum, near the base of the chromosphere \cite[]{val76}.
Procyon's overall warmer photosphere (\teff\ = 6540 K vs. \teff\ =
5770 K) will contribute more flux to this spectral region than does
the Sun.  Furthermore, the warmer components of Procyon's photosphere,
which predominately contribute to the continuum below 160 nm, have a
silicon bound-free opacity approximately three times weaker than
expected for solar photospheric conditions \cite[fig. 6]{tm68}.  Thus
we expect Procyon's continuum between 136 nm and 160 nm to form at
depths beneath its chromosphere.  

The ultraviolet SED longward of 160 nm, where the multicomponent
temperature effects are less prominent, may be impacted by non-LTE
treatment of iron-group elements.  \cite{sh05} find non-LTE models for
the Sun have substantially (up to 20\%) more near-UV flux relative to
LTE models.  Whether such differences would exist for a similar
non-LTE model of Procyon is not clear.

\begin{figure}
\includegraphics[scale=0.8,angle=90]{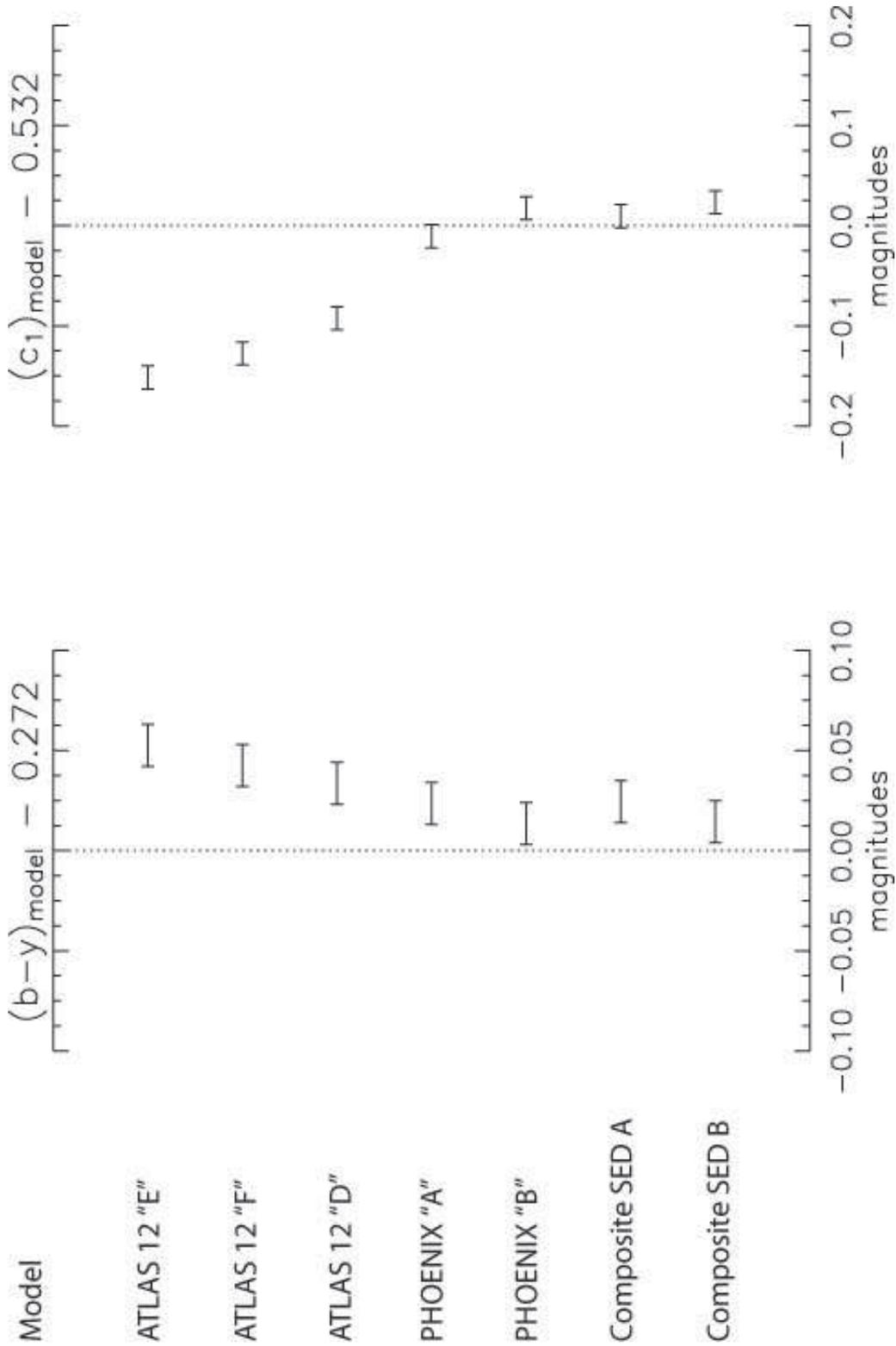}
\caption{A comparison of the synthetic
$(b-y)$ and $c_1$ Str\"{o}mgren indices
from Table \ref{tab:colors} relative to
the observed indices \cite[]{cb70}.}
\label{fig:comp_colors}
\end{figure}

The lower panel of Figure \ref{fig:sed_comparison} shows the {\tt
ATLAS 12} model ``E'' SED matches the Composite SED B model fairly
well in the $u$- and $y$-bands, but has significantly lower flux in
the $b$- and $v$-bands.  In the composite SED model the peaks of the
12 components all contribute flux to the $v$- and $b$-bands, while at
longer wavelengths the flux contrast is lower.  Table \ref{tab:colors}
and Figure \ref{fig:comp_colors} show that the \teff\ = 6530 K {\tt
ATLAS 12} models all predict $(b-y)$ values significantly redder than
observed.  In particular, the $(b-y)$ value for {\tt ATLAS 12} model
``F'', the best fitting 1-D model to the interferometric data, differs
by +0.04 mag, significantly in excess of the estimated uncertainties
(0.012 mag, see \S\ref{zero-points}).  

\begin{figure}
\includegraphics[scale=0.5,angle=0]{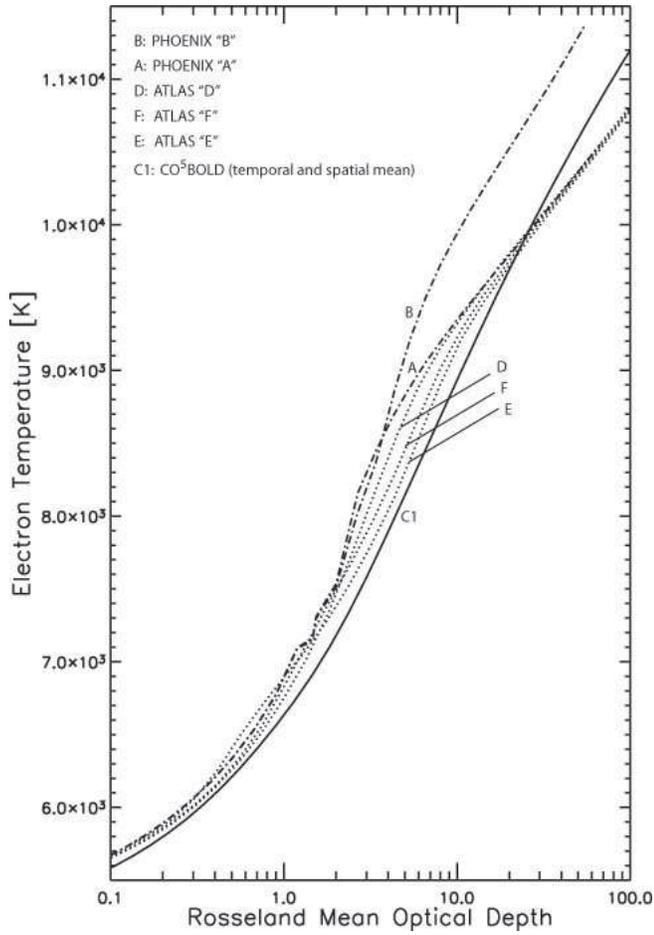}
\caption{A comparison of six model atmosphere temperature on a
mean Rosseland optical depth scale: {\tt PHOENIX} models ``A'' and
``B'' (dash-dot), {\tt ATLAS 12} models ``D'', ``E'', ``F'' (dot), and
{\tt CO$^5$BOLD} model ``C1'' (solid).  The degree of
limb darkening, particularly at 500 nm where the opacity is relatively
low compared to 800 nm and where the intensity contrast as a function
of temperature is relatively high compared to 2.2 \micron, is
sensitive to different convection treatments which affect the
temperature gradient in the interval $\tau_{\rm Ross} = $ 1 to
$\sim3$.  The model with the steepest gradient, {\tt PHOENIX} ``B''
with $\alpha$ = 0.5 and no overshooting, has the strongest limb
darkening.  {\tt ATLAS 12} model ``E'' with 100\% overshooting has the
shallowest gradient and the weakest limb darkening.  {\tt ATLAS 12}
model ``F'' and {\tt CO$^5$BOLD} model ``C1'' have
similar gradients and provide an intermediate degree of limb darkening
which is consistent with the interferometric observations.}
\label{fig:comp_structures}
\end{figure}

The differences in the derived indices between {\tt PHOENIX} model
``A'' and {\tt ATLAS 12} model ``D'', particularly for $c_1$, result
from differences in these models' temperature structures (see
Figure~\ref{fig:comp_structures}).  While these two 1-D models were
constructed to have the same input parameters, they differ in geometry
(spherical vs. plane parallel) and line formation (non-LTE versus LTE
treatment of hydrogen and helium and constant vs. depth-dependent
microturbulence) assumptions.  To investigate this discrepancy, we
converged two LTE, plane-parallel {\tt PHOENIX} models with constant
microturbulences of 0 and 2 \kms.  These models yield colors nearly
identical to model ``A'' , suggesting that the 1-D mixing-length
convection treatments (without overshooting) differ between {\tt
PHOENIX} and {\tt ATLAS 12}.  A difference in the convection
treatments of these two codes is noted by \cite{sh05}.

\section{Discussion}
\label{discussion}
\subsection{Model Differences, Interferometric Uncertainties, and  Future Measurements}
As predicted by AP02, we find, based on comparisons with
high-precision multiwavelength interferometric data, that 1-D model
temperature structures based on standard mixing-length theory (without
overshooting) produce center-to-limb intensity profiles which are too
limb darkened at blue wavelengths relative to near-IR wavelengths.
The magnitude of this effect appears to be slightly smaller than
predicted, $0.4 - 1.5$\% versus 1.6\%, depending on which 1-D model is
adopted.  Furthermore, limb-darkening predictions from a 3-D
hydrodynamical model appear to be consistent with these data, also in
line with AP02's predictions.

Considering the three {\tt ATLAS 12} models, the fit angular diameters
differ negligibly (0.2\%) at 2.2 \micron, but differ by 1.1\% at 500
nm due to different convection treatments.  Considering all seven model
atmospheres, from {\tt PHOENIX}, {\tt ATLAS 12} and {\tt CO$^5$BOLD +
PHOENIX}, the range of fit angular diameters is again small (0.3\%) at
2.2 \micron, with differences up to 1.9\% at 500 nm.  The error bars
on the interferometric data sets yield formal uncertainties of 0.1\%
to 0.2\% in the absolute angular diameters. It should therefore be
possible to choose between these models in the absence of significant
systematic errors.  Any systematic errors in the VLTI/VINCI angular
diameters is expected to be less than 0.3\% \cite[]{procyon_k04}.  The
systematic errors in the Mark III are not as well characterized,
however the three 1-D models without overshooting yield larger angular
diameters at 500 nm relative to 800 nm as expected, while the 3-D {\tt
CO$^5$BOLD} + {\tt PHOENIX} model ``C2'' and {\tt ATLAS 12} model ``F''
for 50\% overshooting yield the same angular diameters.  These
relationships hold out to 2.2 \micron, where model ``C2'' yields the
same angular diameter within 1.3$\sigma$.  This suggests that
systematic errors between the VLTI/VINCI and Mark III data sets are
minimal.

Table~\ref{tab:AD_fits} and Figure~\ref{fig:ad_diffs} show that an
angular diameter derived from the VLTI/VINCI visibilities is the least
sensitive to which model is employed.  The mean angular diameter value
at 2.2 \micron\ together with 1$\sigma$ range in angular diameters
(0.26\%) and the upper limit on the systematic error in the VLTI/VINCI
diameter (0.3\%) yields a best estimate for the angular diameter of
Procyon: 5.404$\pm$0.031 mas. The corresponding best estimates for
\teff, radius, and surface gravity are listed in Table
~\ref{tab:fund_params}.

Interferometric observations with superior calibration are certainly
warranted to check these results.  Independent multiwavelength
angular diameter measurements are in order, however this is not the only
approach presently available.  Limb darkening, in addition to its
affect on these angular diameter measurements, can be constrained by a
precise measurement of the second lobe of the visibility curve.  While
angular diameters derived from the 3-D and standard 1-D models differ
by only 1\% at 500 nm, the squared visibility at the peak of the
second lobe is predicted to differ by 10\% at 500 nm.  At longer
wavelengths the predicted difference is smaller: 6\% at 800 nm and 3\%
at 2.2 \micron.  Precise second lobe measurements at 500 nm and 800 nm
could possibly be made with the Navy Prototype Optical Interferometer
(NPOI) and the Sydney University Stellar Interferometer (SUSI).
Distinguishing between the 3-D {\tt CO$^5$BOLD} model and 1-D
overshooting {\tt ATLAS 12} model ``F'' on the basis of second lobe
measurements would appear to be much more challenging.  The squared
visibility amplitudes at the peak of the 2nd lobe for these two models
differ by less than 1\% at 500 nm, 800 nm, and 2.2 \micron.

\subsection{Procyon's Mean Temperature Gradient}
\label{gradient}
The continuum forming region of Procyon's atmosphere is expected to have
a physical thickness which is very small ($<0.1$\%) in comparison with
its stellar radius.  Therefore, the wavelength-dependent uniform disk
angular diameter is unlikely to result from an extended atmosphere where
significantly different physical depths are probed at different
wavelengths, as in the case of cool supergiants.  Instead this
effect must reside in the distribution of temperature with optical
depth.

The best fit angular diameter values presented in Table
~\ref{tab:AD_fits} and Figure ~\ref{fig:ad_diffs} are quite
insensitive to the model effective temperature. Fits obtained from a
grid of {\tt PHOENIX} models spanning a 2000 K range yield angular
diameter differences of only 0.007 mas, 0.006 mas 0.004 mas at 500 nm,
800 nm, and 2.2 \micron\ respectively; values comparable to the formal
fit uncertainties.  The atmosphere models in Table ~\ref{tab:AD_fits}
yield different angular diameters because they have different
temperature gradients, as shown in Figure \ref{fig:comp_structures}.

\begin{figure}
\includegraphics[scale=0.7,angle=0]{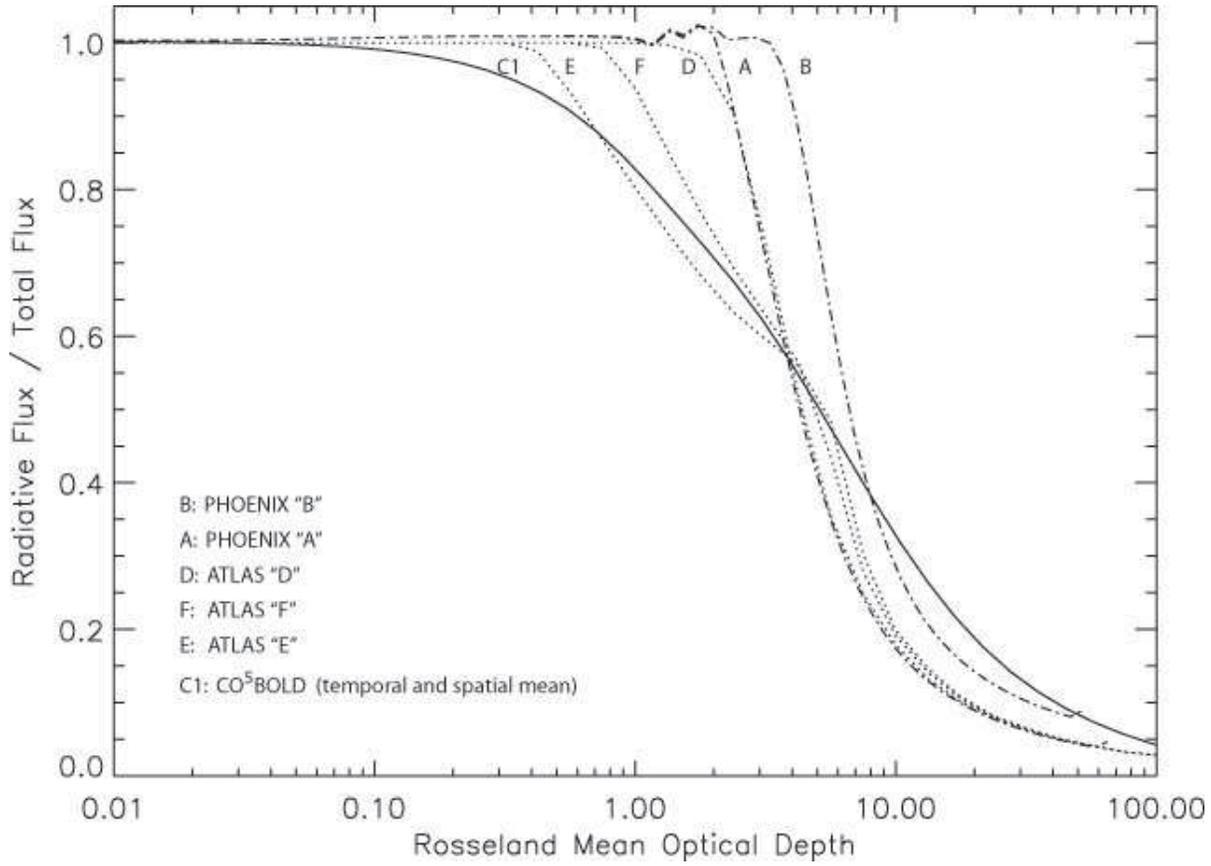}
\caption{A comparison of the six model atmosphere radiative flux
structures on a mean Rosseland optical depth scale: {\tt PHOENIX} models
``A'' and ``B'' (dash-dot), {\tt ATLAS 12} models ``D'', ``E'', ``F''
(dot), and {\tt CO$^5$BOLD} model ``C1'' (solid).  The 1-D
overshooting models ``E'' and ``F'' and the temporally and spatially
averaged 3-D model ``C1'' all have a significant fraction (5\% to
15\%) of their total flux in convection at an optical depth of unity,
while the other 1-D models are fully radiative at this depth.}
\label{fig:comp_radiative_flux}
\end{figure}

The mean temperature structure of the 3-D {\tt CO$^5$BOLD} model and
the 1-D {\tt ATLAS 12} model ``F'' have shallower temperature
gradients in the range $ 1 \leq \tau_{\rm ross} \leq 5$ relative to
the 1-D models without overshooting.  This is expected because the
radiative flux should be reduced when the convective flux is
increased.  Figure \ref{fig:comp_radiative_flux} shows the fraction of
radiative flux as a function of optical depth for these models.  The
mean 3-D model and the 1-D model with overshooting have only $\sim
80$\% of their total flux in radiation at $\tau_{\rm ross} = 1$.  At
depths $\tau_{\rm ross} \lesssim 8$ the mean 3-D and the 1-D
overshooting radiative flux structures are quite similar.  These best
fitting models to the interferometric data show significant convective
flux at $\tau_{\rm ross} \lesssim 1$ and we conclude that the
interferometric data provide evidence for convective
overshooting in Procyon's atmosphere.

The shallower temperature gradient introduced by overshooting reduces
the relative limb darkening between 500 nm and 2.2 \micron.  This
accounts for the variation in the best fit $\theta_{\rm LD}$ values
for models with different convection treatments.  An analysis of the
contribution functions, $S(\tau_{\lambda}) e^{-\tau_{\lambda}}$, at
the three wavelengths shows the 500 nm continuum forms in a range of
depths partially below the depths forming the 800 nm continuum.  The
800 nm band lies near the peak of the bound-free H$^-$ opacity, the
dominant continuum opacity in the optical.  At 2.2 \micron\ the H$^-$
opacity, here free-free beyond the 1.6 \micron\ H$^-$ opacity minimum,
is less than at 800 nm and similar to the 500 nm opacity.  However, at
2.2 \micron\ the intensity is much less sensitive to the temperature of
Planckian radiation than at 500 nm.  For the Sun, it was pointed out
by \cite{HSRA71} that the lack of a steeply rising Planck function
(with temperature) at 1.6 \micron\ means that the intensities emerging
from the hot convective layers are better probed at wavelengths
immediately to the red of the Balmer limit (365 nm) even if the
optical depth is a factor of two larger there than at the 1.6 \micron\
opacity minimum.  In short, similar depths are probed by the 500 nm
and 2.2 \micron\ bands, but the radiation field is much more sensitive
the temperature gradient at 500 nm compared to 2.2 \micron.

One-dimensional models without overshooting cannot be made more
consistent with the multiwavelength visibility data by adjusting the
mixing length parameter $\alpha$.  Based on the two stand-alone {\tt
PHOENIX} models ``A'' and ``B'' (see Table \ref{tab:AD_fits}),
changing $\alpha$ from 1.25 to 0.5 has no effect on the derived
angular diameters at 800 nm and 2.2 \micron. At 500 nm the two best
fit angular diameters differ by 0.35\%.  Note that model ``B'' with
$\alpha = 0.5$ yields a slightly larger angular diameter at 500 nm
relative to the $\alpha = 1.25$ model.  Thus, reducing $\alpha$
slightly increases the degree of limb darkening for a model which is
already too limb darkened compared to the observations presented here.

\subsection{Photometry and 1-D Overshooting}
As discussed above, we find that the 1-D {\tt ATLAS 12} model ``F''
with 50\% overshooting fits the visibility data as well as the 3-D
{\tt CO$^5$BOLD} model and that the two models have similar mean
temperature gradients.  However, models with mixing-length theory
convection and overshooting have been subject to extensive comparisons
with photometry, with the results favoring models without overshooting
\cite[]{cgk97,sk97}.  In general, 1-D convection treatments
(e.g. mixing-length variants, turbulent convection) for model stellar
atmospheres with otherwise identical parameters yield different SEDs.
Str\"{o}mgren photometric indices are often employed to facilitate the
comparison of these different synthetic SEDs with a large number of
real SEDs.  Such studies \cite[][and references therein]{heiter02}
generally find that 1-D atmosphere models with no overshooting and
$\alpha = 0.5$ more closely reproduce the colors of real atmospheres
relative to other 1-D models.  This is demonstrated for Procyon by
\cite{sk97} who show that to reproduce the observed $(b-y)$ index with
an {\tt ATLAS 9} overshooting model requires a model with \teff\ =
6830$\pm$ 142 K (the uncertainly based on photometry alone), a value
in excess of their derived interferometric value, \teff\ = 6560 $\pm$
130 K.  This confirms the findings of \cite{cgk97}.  These results
appear to be strengthened by our more precise value for the
interferometric \teff\ value, now 6543$\pm$87 K, based on the mean
K-band angular diameter value (see Table~\ref{tab:fund_params}).  Now
the best fit 1-D overshooting model to the photometry would appear to
require a model \teff\ value even more inconsistent with the
interferometric \teff.

\cite{heiter02} find that reducing the mixing-length parameter
$\alpha$ in {\tt ATLAS 9} models yields synthetic indices more
consistent with observations.  Our comparisons suggest that 1-D models
with lower $\alpha$ values better match photometry because this better
approximates the composite SED in the $b$- and $v$-bands.  As a
result, constraining the 1-D $\alpha$ parameter with Str\"{o}mgren
photometry most likely constrains not the temperature structure of the
1-D model, but instead reflects the lack of hotter temperature
components in the 1-D model, the same components responsible for the
continuum below 160 nm.

The composite SED models reproduce photometric indices quite close to
the observed values and match fairly well the full UV continuum flux
distribution.  On the basis of the spatially and temporally averaged
3-D radiative flux structure (Figure \ref{fig:comp_radiative_flux})
one might guess a much redder $(b-y)$ color for this model because of
the similarly to {\tt ATLAS 12} model ``F''.  However, unlike the 1-D
case, in the 3-D case the distribution of surface temperatures
suppresses the one-to-one connection between the mean photospheric SED
and the mean temperature structure.  Both composite models predict
$(b-y)$ values slightly redder than observed, but much more consistent
than the 1-D overshooting models.  A slightly warmer {\tt CO$^5$BOLD}
model (\teff\ = 6550 K instead of 6500 K), more consistent with the
interferometric value 6543 K, is expected to yield a slightly warmer
distribution of temperatures and hence a bluer $(b-y)$.  A more
consistent modeling approach, where the colors are computed directly
from the 3-D model intensities, should clearly be pursued once the
wavelength-dependent opacities in 3-D models improve.

\section{Summary}
\label{summary}
We find that the multiwavelength visibility measurements of
Procyon obtained by the Mark III and the VLTI interferometers
characterize the mean temperature structure of its atmosphere while UV
and optical spectrophotometry characterize the spatial distribution of
surface temperatures expected from granulation on its surface.  We
find that predictions from a three-dimensional {\tt CO$^5$BOLD}
hydrodynamic model atmosphere of Procyon match the interferometric,
spectrophotometric, and photometric observations simultaneously,
providing confidence in 3-D model predictions which until now had been
tested predominantly with high-resolution spectroscopy.  Based on 3-D
and 1-D model comparisons, the interferometric data are consistent
with a temperature structure with significant convective overshooting.
While a 1-D model atmosphere with some overshooting matches the
interferometric data, all 1-D models fail to match Procyon's spectral
energy distribution, most evidently below 160 nm.  Str\"{o}mgren
photometric indices appear to be sensitive to the multicomponent
brightness temperature distribution on Procyon's surface.  This would
seem to severely complicate conclusions regarding convection in
similar stars based on 1-D model atmosphere comparisons to optical
spectrophotometric data.

\begin{acknowledgments}
D. Mozurkewich kindly provided the Mark III visibility data.  We thank
R. L. Kurucz for computing the {\tt ATLAS 12} spectra and radiation
fields and for critical comments.  These data and models were vital for
this project.  Thanks to S. T. Ridgway and S. N. Shore for their
careful readings of the manuscript and for helpful discussions.
I. Hubeny kindly provided critical comments.  Thanks to A. M\'{e}rand
for many helpful discussions and correspondence. Thanks to our
anonymous referee for comments regarding the interferometric
calibrations and for improving the organization of the paper.  Thanks
to the {\tt PHOENIX} development team for their support and interest
in this work.  This work was performed in part under contract with the
Jet Propulsion Laboratory (JPL) funded by NASA through the Michelson
Fellowship Program.  JPL is managed for NASA by the California
Institute of Technology.  NOAO is operated by AURA, Inc, under
cooperative agreement with the National Science Foundation. HGL
acknowledges financial support by the Swedish Research Council and the
Royal Physiographic Society in Lund.  The VLTI is operated by the
European Southern Observatory at Cerro Paranal, Chile.  The Mark III
was funded by the Office of Naval Research and the Oceanographer of
the Navy.  This research has made use of NASA's Astrophysics Data
System, SIMBAD database, operated at CDS, Strasbourg, France. Some of
the data presented in this paper was obtained from the Multimission
Archive at the Space Telescope Science Institute (MAST). STScI is
operated by the Association of Universities for Research in Astronomy,
Inc., under NASA contract NAS5-26555. Support for MAST for non-HST
data is provided by the NASA Office of Space Science via grant
NAG5-7584 and by other grants and contracts.
\end{acknowledgments}

\clearpage
\LongTables
\begin{deluxetable}{rrrrrrrr}
\tablecolumns{8} 
\tabletypesize{\footnotesize}
\tablecaption{VLTI/VINCI K-band Visibility Measurements}
\tablewidth{0pt}
\tablehead{ 
\colhead{Julian Date}
&\colhead{Projected} 
&\colhead{Position} 
&\colhead{$V^2$}
&\colhead{$\sigma V^2_{\rm stat}$}
&\colhead{$\sigma V^2_{\rm sys}$}
&\colhead{$\sigma V^2_{\rm total}$}
&\colhead{Calibration Star}\\
\colhead{}
&\colhead{Baseline}
&\colhead{Angle}
&\colhead{$\times$ 100}
&\colhead{$\times$ 100}
&\colhead{$\times$ 100}
&\colhead{$\times$ 100}
&\colhead{}\\
\colhead{}
&\colhead{(meters)}
&\colhead{(degrees)}
&\colhead{}
&\colhead{}
&\colhead{}
&\colhead{}
&\colhead{}
}
\startdata 
2452674.59969  &21.900  &73.45   &83.01  &1.90   &0.12   &1.90   &$\alpha$\,CMa \\
2452674.60482  &22.192  &73.51   &83.50  &2.20   &0.12   &2.21   &$\alpha$\,CMa \\
2452674.60889  &22.409  &73.54   &83.76  &2.18   &0.12   &2.18   &$\alpha$\,CMa \\
2452672.72358  &23.005  &70.29   &81.90  &1.31   &0.08   &1.31   &$\alpha$\,CMa \\
2452682.69435  &23.081  &70.42   &86.17  &3.20   &0.07   &3.20   &$\alpha$\,CMa \\
2452685.68579  &23.095  &70.45   &84.07  &1.86   &0.08   &1.86   &$\alpha$\,CMa \\
2452671.72379  &23.104  &70.46   &84.09  &3.61   &0.07   &3.61   &$\alpha$\,CMa \\
2452681.69611  &23.118  &70.49   &84.93  &3.86   &0.08   &3.87   &$\alpha$\,CMa \\
2452672.72041  &23.129  &70.50   &83.48  &1.13   &0.08   &1.14   &$\alpha$\,CMa \\
2452683.68954  &23.160  &70.56   &84.70  &2.88   &0.09   &2.88   &$\alpha$\,CMa \\
2452682.69041  &23.227  &70.68   &83.64  &3.12   &0.07   &3.12   &$\alpha$\,CMa \\
2452681.69192  &23.271  &70.76   &83.88  &3.81   &0.08   &3.81   &$\alpha$\,CMa \\
2452679.69711  &23.280  &70.77   &81.32  &3.88   &0.23   &3.89   &$\alpha$\,CMa \\
2452671.71859  &23.293  &70.80   &82.83  &3.40   &0.07   &3.40   &$\alpha$\,CMa \\
2452683.68469  &23.331  &70.87   &82.44  &2.73   &0.09   &2.73   &$\alpha$\,CMa \\
2452682.68572  &23.386  &70.97   &80.90  &3.01   &0.07   &3.01   &$\alpha$\,CMa \\
2452679.69316  &23.410  &71.01   &83.33  &2.38   &0.23   &2.39   &$\alpha$\,CMa \\
2452671.71484  &23.415  &71.02   &82.49  &2.19   &0.07   &2.20   &$\alpha$\,CMa \\
2452683.68116  &23.443  &71.08   &82.54  &2.71   &0.09   &2.71   &$\alpha$\,CMa \\
2452671.70658  &23.644  &71.48   &80.91  &2.43   &0.07   &2.43   &$\alpha$\,CMa \\
2452671.70225  &23.741  &71.71   &81.93  &2.43   &0.07   &2.43   &$\alpha$\,CMa \\
2452684.66566  &23.764  &71.76   &83.54  &2.51   &0.09   &2.51   &$\alpha$\,CMa \\
2452685.66048  &23.810  &71.88   &83.32  &1.83   &0.09   &1.84   &$\alpha$\,CMa \\
2452684.62226  &23.819  &73.24   &80.76  &2.43   &0.08   &2.43   &$\alpha$\,CMa \\
2452671.69791  &23.824  &71.91   &85.10  &2.57   &0.07   &2.57   &$\alpha$\,CMa \\
2452684.66129  &23.842  &71.96   &82.11  &2.47   &0.08   &2.47   &$\alpha$\,CMa \\
2452685.65611  &23.879  &72.07   &82.65  &1.71   &0.09   &1.72   &$\alpha$\,CMa \\
2452683.62916  &23.884  &73.15   &82.34  &2.74   &0.09   &2.74   &$\alpha$\,CMa \\
2452684.62693  &23.891  &73.14   &84.20  &2.66   &0.09   &2.66   &$\alpha$\,CMa \\
2452682.66269  &23.901  &72.15   &80.19  &2.99   &0.07   &2.99   &$\alpha$\,CMa \\
2452684.65703  &23.903  &72.15   &82.22  &2.47   &0.08   &2.47   &$\alpha$\,CMa \\
2452671.69151  &23.915  &72.20   &82.27  &2.16   &0.07   &2.16   &$\alpha$\,CMa \\
2452685.65156  &23.934  &72.27   &80.65  &1.68   &0.09   &1.68   &$\alpha$\,CMa \\
2452683.63367  &23.938  &73.04   &82.01  &2.76   &0.09   &2.76   &$\alpha$\,CMa \\
2452684.63099  &23.938  &73.04   &80.15  &2.72   &0.08   &2.72   &$\alpha$\,CMa \\
2452682.65863  &23.945  &72.31   &82.83  &3.08   &0.07   &3.08   &$\alpha$\,CMa \\
2452672.68439  &23.958  &72.37   &81.00  &1.13   &0.08   &1.13   &$\alpha$\,CMa \\
2452674.67842  &23.962  &72.39   &81.05  &1.80   &0.10   &1.81   &$\alpha$\,CMa \\
2452683.65385  &23.962  &72.39   &83.33  &2.78   &0.09   &2.78   &$\alpha$\,CMa \\
2452683.63773  &23.971  &72.93   &81.89  &2.80   &0.09   &2.80   &$\alpha$\,CMa \\
2452682.65360  &23.981  &72.50   &80.45  &3.00   &0.07   &3.00   &$\alpha$\,CMa \\
2452684.64797  &23.981  &72.51   &82.02  &2.47   &0.08   &2.48   &$\alpha$\,CMa \\
2452685.63493  &23.985  &72.85   &82.08  &1.94   &0.09   &1.94   &$\alpha$\,CMa \\
2452683.64960  &23.986  &72.55   &83.19  &2.78   &0.09   &2.78   &$\alpha$\,CMa \\
2452674.67396  &23.987  &72.56   &81.92  &1.82   &0.10   &1.82   &$\alpha$\,CMa \\
2452672.67922  &23.988  &72.56   &79.78  &1.16   &0.08   &1.16   &$\alpha$\,CMa \\
2452685.64373  &23.988  &72.56   &79.08  &1.91   &0.08   &1.91   &$\alpha$\,CMa \\
2452684.63971  &23.991  &72.79   &81.00  &2.43   &0.08   &2.44   &$\alpha$\,CMa \\
2452684.64410  &23.993  &72.65   &81.56  &2.45   &0.08   &2.45   &$\alpha$\,CMa \\
2452672.67394  &23.994  &72.74   &80.85  &1.10   &0.08   &1.10   &$\alpha$\,CMa \\
2452674.67012  &23.995  &72.69   &81.18  &1.80   &0.10   &1.80   &$\alpha$\,CMa \\
2452683.64541  &23.995  &72.69   &81.71  &2.72   &0.09   &2.72   &$\alpha$\,CMa \\
2452685.63998  &23.995  &72.69   &79.22  &1.68   &0.08   &1.68   &$\alpha$\,CMa \\
2453002.63052  &42.518  &69.88   &54.52  &1.23   &0.22   &1.25   &30\,Gem \\
2453002.63564  &43.943  &70.39   &51.89  &1.16   &0.21   &1.18   &30\,Gem \\
2453002.66192  &50.648  &72.27   &40.22  &0.51   &0.18   &0.54   &30\,Gem \\
2453002.66734  &51.889  &72.53   &36.85  &0.71   &0.16   &0.73   &30\,Gem \\
2452339.59761  &55.734  &147.99  &32.92  &1.33   &0.44   &1.40   &$\alpha$\,CMa \\
2452339.59331  &56.237  &147.32  &29.39  &0.99   &0.40   &1.07   &$\alpha$\,CMa \\
2452339.58907  &56.731  &146.69  &31.28  &1.11   &0.42   &1.18   &$\alpha$\,CMa \\
2453002.69427  &57.239  &73.37   &28.79  &0.50   &0.13   &0.52   &30\,Gem \\
2453029.62077  &57.277  &73.37   &29.14  &2.01   &0.10   &2.02   &18\,Mon \\
2453002.69950  &58.111  &73.45   &27.98  &0.48   &0.13   &0.50   &30\,Gem \\
2453029.62595  &58.137  &73.45   &27.79  &1.92   &0.10   &1.92   &18\,Mon \\
2452994.73707  &60.383  &73.58   &25.89  &0.45   &0.15   &0.47   &HR\,1799 \\
2453002.72628  &61.657  &73.57   &22.37  &0.73   &0.09   &0.73   &18\,Mon \\
2453002.73158  &62.172  &73.54   &22.10  &1.05   &0.09   &1.06   &18\,Mon \\
2452996.75146  &62.477  &73.51   &22.06  &0.25   &0.08   &0.27   &6\,Leo \\
2452996.75644  &62.863  &73.45   &21.48  &0.26   &0.08   &0.27   &6\,Leo \\
2452995.76035  &62.946  &73.43   &21.06  &0.33   &0.08   &0.34   &6\,Leo \\
2452996.81799  &62.952  &71.42   &21.44  &0.33   &0.11   &0.34   &6\,Leo \\
2452995.81464  &63.326  &71.74   &20.80  &0.27   &0.10   &0.28   &6\,Leo \\
2452996.81167  &63.339  &71.75   &20.41  &0.27   &0.10   &0.29   &6\,Leo \\
2452995.80179  &63.845  &72.31   &20.26  &0.27   &0.10   &0.29   &6\,Leo \\
2452996.78120  &63.942  &72.93   &19.80  &0.18   &0.08   &0.20   &6\,Leo \\
2453002.76593  &63.957  &72.89   &20.05  &0.59   &0.08   &0.60   &18\,Mon \\
2452996.78597  &63.988  &72.78   &19.85  &0.20   &0.08   &0.22   &6\,Leo \\
2453002.77184  &63.991  &72.71   &19.88  &0.66   &0.08   &0.66   &18\,Mon \\
\enddata
\label{tab:vlti_data}
\end{deluxetable}
\begin{deluxetable}{rrrrr}
\tablecolumns{5} 
\tabletypesize{\footnotesize}
\tablecaption{Mark III 500 nm Visibility Measurements}
\tablewidth{0pt}
\tablehead{ 
\colhead{Julian Date} 
&\colhead{Projected} 
&\colhead{Position} 
&\colhead{$V^2$}
&\colhead{$\sigma V^2_{\rm total}$}\\
\colhead{}
&\colhead{Baseline}
&\colhead{Angle}
&\colhead{}
&\colhead{}\\
\colhead{}
&\colhead{(meters)}
&\colhead{(degrees)}
&\colhead{}
&\colhead{}
}
\startdata              

2447452.956    &2.803  & 23.049   & 1.024   & 0.106\\
2447452.970    &2.768  & 20.864   & 0.900   & 0.089\\
2447452.986    &2.728  & 17.918   & 0.944   & 0.072\\
2447453.004    &2.689  & 14.543   & 0.966   & 0.073\\
2447453.041    &2.630  &  6.448   & 1.013   & 0.075\\
2447453.924    &2.877  & 27.258   & 0.859   & 0.058\\
2447453.936    &2.847  & 25.646   & 0.911   & 0.049\\
2447453.949    &2.814  & 23.750   & 0.984   & 0.051\\
2447453.962    &2.781  & 21.688   & 0.904   & 0.046\\
2447453.981    &2.734  & 18.425   & 0.940   & 0.048\\
2447454.921    &5.093  & 27.288   & 0.894   & 0.053\\
2447454.957    &4.934  & 22.135   & 0.780   & 0.046\\
2447454.969    &4.878  & 20.047   & 0.804   & 0.048\\
2447454.980    &4.833  & 18.154   & 0.884   & 0.051\\
2447454.993    &4.778  & 15.552   & 0.933   & 0.054\\
2447455.003    &4.741  & 13.490   & 0.925   & 0.054\\
2447455.015    &4.705  & 11.116   & 0.927   & 0.054\\
2447455.024    &4.680  &  9.063   & 0.954   & 0.055\\
2447455.037    &4.653  &  6.184   & 0.831   & 0.051\\
2447455.050    &4.636  &  3.183   & 0.893   & 0.052\\
2447455.910    &6.686  & 28.266   & 0.578   & 0.048\\
2447455.928    &6.587  & 25.979   & 0.730   & 0.057\\
2447455.945    &6.490  & 23.593   & 0.659   & 0.052\\
2447455.962    &6.384  & 20.683   & 0.762   & 0.059\\
2447455.974    &6.319  & 18.661   & 0.694   & 0.053\\
2447455.985    &6.258  & 16.546   & 0.755   & 0.057\\
2447455.996    &6.204  & 14.363   & 0.802   & 0.061\\
2447456.007    &6.155  & 12.079   & 0.828   & 0.063\\
2447456.018    &6.114  &  9.724   & 0.836   & 0.064\\
2447456.030    &6.079  &  7.157   & 0.826   & 0.063\\
2447456.041    &6.055  &  4.478   & 0.800   & 0.061\\
2447456.916   &22.193  & 27.143   & 0.010   & 0.002\\
2447456.935   &21.836  & 24.598   & 0.016   & 0.003\\
2447457.046   &20.215  &  2.793   & 0.035   & 0.009\\
2447457.964   &21.220  & 19.514   & 0.020   & 0.002\\
2447457.976   &20.984  & 17.152   & 0.024   & 0.003\\
2447457.993   &20.703  & 13.805   & 0.031   & 0.002\\
2447458.008   &20.505  & 10.815   & 0.030   & 0.002\\
2447458.019   &20.377  &  8.311   & 0.032   & 0.002\\
2447458.034   &20.256  &  4.843   & 0.033   & 0.002\\
2447458.044   &20.213  &  2.678   & 0.038   & 0.002\\
2447458.053   &20.195  &  0.435   & 0.040   & 0.002\\
2447460.952   &29.068  & 20.149   & 0.006   & 0.002\\
2447460.970   &28.613  & 16.858   & 0.003   & 0.002\\
\enddata
\label{tab:m3_500_data}
\end{deluxetable}

\clearpage

\begin{deluxetable}{rrrrr}
\tablecolumns{5} 
\tabletypesize{\footnotesize}
\tablecaption{Mark III 800 nm Visibility Measurements}
\tablewidth{0pt}
\tablehead{ 
\colhead{Julian Date} 
&\colhead{Projected} 
&\colhead{Position} 
&\colhead{$V^2$}
&\colhead{$\sigma V^2_{\rm total}$}\\
\colhead{}
&\colhead{Baseline}
&\colhead{Angle}
&\colhead{}
&\colhead{}\\
\colhead{}
&\colhead{(meters)}
&\colhead{(degrees)}
&\colhead{}
&\colhead{}
}
\startdata              
2447452.956   & 2.803 &  23.048  &  1.016  &  0.049\\
2447452.970   & 2.768 &  20.863  &  0.948  &  0.043\\
2447452.986   & 2.728 &  17.918  &  1.009  &  0.037\\
2447453.004   & 2.689 &  14.544  &  0.988  &  0.035\\
2447453.022   & 2.656 &  10.718  &  0.997  &  0.035\\
2447453.041   & 2.630 &   6.447  &  1.015  &  0.036\\
2447453.924   & 2.877 &  27.260  &  0.964  &  0.042\\
2447453.936   & 2.847 &  25.647  &  0.968  &  0.029\\
2447453.949   & 2.814 &  23.750  &  1.003  &  0.030\\
2447453.962   & 2.781 &  21.688  &  0.963  &  0.028\\
2447453.981   & 2.734 &  18.423  &  0.970  &  0.027\\
2447454.921   & 5.093 &  27.288  &  0.964  &  0.027\\
2447454.957   & 4.934 &  22.135  &  0.908  &  0.026\\
2447454.969   & 4.878 &  20.049  &  0.919  &  0.026\\
2447454.980   & 4.833 &  18.155  &  0.967  &  0.026\\
2447454.993   & 4.778 &  15.552  &  0.986  &  0.027\\
2447455.003   & 4.741 &  13.491  &  0.978  &  0.027\\
2447455.015   & 4.705 &  11.115  &  0.972  &  0.027\\
2447455.024   & 4.680 &   9.062  &  0.977  &  0.026\\
2447455.037   & 4.653 &   6.185  &  0.933  &  0.032\\
2447455.050   & 4.636 &   3.183  &  0.960  &  0.027\\
2447455.928   & 6.587 &  25.980  &  0.881  &  0.030\\
2447455.945   & 6.490 &  23.593  &  0.845  &  0.030\\
2447455.962   & 6.384 &  20.683  &  0.899  &  0.030\\
2447455.974   & 6.319 &  18.661  &  0.856  &  0.029\\
2447455.985   & 6.258 &  16.546  &  0.900  &  0.030\\
2447455.996   & 6.204 &  14.363  &  0.920  &  0.030\\
2447456.007   & 6.155 &  12.078  &  0.922  &  0.031\\
2447456.018   & 6.114 &   9.724  &  0.941  &  0.031\\
2447456.030   & 6.079 &   7.157  &  0.939  &  0.031\\
2447456.041   & 6.055 &   4.478  &  0.928  &  0.030\\
2447438.947   &18.608 &  29.260  &  0.364  &  0.013\\
2447438.968   &18.309 &  26.817  &  0.380  &  0.012\\
2447438.988   &17.984 &  23.980  &  0.395  &  0.012\\
2447439.024   &17.413 &  17.976  &  0.415  &  0.013\\
2447439.043   &17.146 &  14.289  &  0.438  &  0.014\\
2447456.916   &22.193 &  27.143  &  0.244  &  0.006\\
2447456.935   &21.836 &  24.598  &  0.250  &  0.006\\
2447457.046   &20.215 &   2.793  &  0.320  &  0.016\\
2447457.964   &21.220 &  19.514  &  0.271  &  0.006\\
2447457.976   &20.984 &  17.152  &  0.290  &  0.007\\
2447457.993   &20.703 &  13.805  &  0.306  &  0.006\\
2447458.008   &20.505 &  10.815  &  0.312  &  0.006\\
2447458.019   &20.377 &   8.311  &  0.307  &  0.007\\
2447458.034   &20.256 &   4.843  &  0.335  &  0.007\\
2447458.044   &20.213 &   2.678  &  0.332  &  0.007\\
2447458.053   &20.195 &   0.435  &  0.335  &  0.007\\
2447460.952   &29.068 &  20.149  &  0.068  &  0.003\\
2447460.970   &28.613 &  16.858  &  0.079  &  0.004\\
2447478.899   &26.960 &  20.833  &  0.108  &  0.007\\
2447478.935   &26.154 &  14.128  &  0.125  &  0.009\\
2447478.950   &25.887 &  11.006  &  0.139  &  0.010\\
2447478.965   &25.679 &   7.698  &  0.149  &  0.010\\
2447484.836   &28.099 &  27.661  &  0.077  &  0.016\\
2447484.867   &27.366 &  23.477  &  0.103  &  0.010\\
2447484.878   &27.077 &  21.631  &  0.121  &  0.015\\
2447484.884   &26.926 &  20.595  &  0.103  &  0.011\\
2447484.891   &26.773 &  19.497  &  0.117  &  0.014\\
2447484.897   &26.630 &  18.398  &  0.120  &  0.017\\
2447484.939   &25.806 &   9.841  &  0.143  &  0.013\\
2447484.951   &25.659 &   7.303  &  0.147  &  0.015\\
2447484.962   &25.552 &   4.594  &  0.146  &  0.014\\
2447484.974   &25.493 &   1.879  &  0.143  &  0.012\\
2447484.981   &25.482 &   0.339  &  0.154  &  0.013\\
2447484.987   &25.486 &  -1.146  &  0.135  &  0.011\\
2447484.993   &25.502 &  -2.497  &  0.124  &  0.016\\
2447833.958   &24.330 &  14.949  &  0.184  &  0.002\\
2447833.974   &24.055 &  11.678  &  0.197  &  0.003\\
2447833.983   &23.923 &   9.706  &  0.209  &  0.003\\
2447833.989   &23.855 &   8.511  &  0.202  &  0.003\\
2447833.997   &23.767 &   6.625  &  0.206  &  0.003\\
2447834.002   &23.722 &   5.442  &  0.209  &  0.003\\
2447834.011   &23.670 &   3.558  &  0.217  &  0.005\\
2447834.024   &23.632 &   0.512  &  0.214  &  0.004\\
2447834.029   &23.633 &  -0.749  &  0.198  &  0.004\\
2447834.036   &23.648 &  -2.353  &  0.208  &  0.003\\
2447834.042   &23.673 &  -3.700  &  0.203  &  0.004\\
2447834.049   &23.715 &  -5.208  &  0.205  &  0.003\\
2447834.055   &23.766 &  -6.598  &  0.207  &  0.006\\
2447834.061   &23.831 &  -8.048  &  0.182  &  0.007\\
\enddata
\label{tab:m3_800_data}
\end{deluxetable}

\begin{deluxetable}{clllrr}
\tablecolumns{6} 
\tabletypesize{\footnotesize}
\tablecaption{Synthetic Str\"{o}mgren Indices for Composite SED Components}
\tablewidth{0pt}
\tablehead{
\colhead{Surface Area Weight}	    
&\colhead{\teff (K)}	    
&\colhead{$\log(g)$}	    
&\colhead{$\alpha$}
&\colhead{$(b-y)$}	    
&\colhead{$c_1$}	\\    
\colhead{}	    
&\colhead{}	    
&\colhead{}	    
&\colhead{}	    
&\colhead{}
&\colhead{}
}
\startdata              
\cutinhead{Composite SED A, $\beta$ = 0.6 linear limb darkening}
0.002  &5483 &3.95 &1.25   &0.452  &0.469    \\ 
0.027  &5693 &3.95 &1.25   &0.416  &0.446    \\ 
0.093  &5893 &3.95 &1.25   &0.384  &0.436    \\ 
0.125  &6074 &3.95 &1.25   &0.356  &0.445    \\ 
0.120  &6241 &3.95 &1.25   &0.332  &0.464    \\ 
0.108  &6395 &3.95 &1.25   &0.311  &0.491    \\ 
0.114  &6538 &3.95 &1.25   &0.291  &0.521    \\ 
0.134  &6672 &3.95 &1.25   &0.274  &0.554    \\ 
0.139  &6799 &3.95 &1.25   &0.257  &0.586    \\ 
0.106  &6920 &3.95 &1.25   &0.242  &0.620    \\ 
0.029  &7034 &3.95 &1.25   &0.227  &0.652    \\ 
0.002  &7145 &3.95 &1.25   &0.213  &0.684    \\ 
\cutinhead{Composite SED B, $\beta$ = 0.5 linear limb darkening}
0.002  &5539 &3.95 &1.25    &0.442  &0.461    \\ 
0.027  &5751 &3.95 &1.25    &0.406  &0.444    \\ 
0.093  &5953 &3.95 &1.25    &0.375  &0.437    \\ 
0.125  &6136 &3.95 &1.25    &0.347  &0.451    \\ 
0.120  &6304 &3.95 &1.25    &0.323  &0.474    \\ 
0.108  &6460 &3.95 &1.25    &0.302  &0.504    \\ 
0.114  &6604 &3.95 &1.25    &0.283  &0.536    \\ 
0.134  &6740 &3.95 &1.25    &0.265  &0.571    \\ 
0.139  &6869 &3.95 &1.25    &0.248  &0.606    \\ 
0.106  &6990 &3.95 &1.25    &0.233  &0.639    \\ 
0.029  &7106 &3.95 &1.25    &0.218  &0.672    \\ 
0.002  &7217 &3.95 &1.25    &0.205  &0.704    \\ 
\enddata 
\label{tab:component_colors}
\end{deluxetable}


\begin{deluxetable}{llrrr}
\tablecolumns{5} 
\tabletypesize{\footnotesize}
\tablecaption{Fit Angular Diameters}
\tablewidth{0pt}
\tablehead{ 
\colhead{Model} 
&\colhead{Parameters}
&\colhead{Mark III}
&\colhead{Mark III}
&\colhead{VLTI/VINCI}\\
\colhead{}
&\colhead{}
&\colhead{500 nm}
&\colhead{800 nm}
&\colhead{2.2\micron}\\
\colhead{}
&\colhead{}
&\colhead{(mas)}
&\colhead{(mas)}
&\colhead{(mas)}
} 
\startdata               
``A''  {\tt PHOENIX}    &\teff=6530 K, $\log(g)=3.95$, $\alpha=1.25$                         &5.449$\pm$0.012   &5.416$\pm$0.005  &5.411$\pm$0.006 \\
``B'' {\tt PHOENIX}     &\teff=6530 K, $\log(g)=3.95$, $\alpha=0.5$                          &5.468$\pm$0.012   &5.415$\pm$0.005  &5.410$\pm$0.006  \\
``C1'' {\tt CO$^5$BOLD} + {\tt PHOENIX} &\teff=6500 K, $\log(g)=4.0$, global mean structure  &5.387$\pm$0.011   &5.388$\pm$0.005  &5.402$\pm$0.006  \\
``C2'' {\tt CO$^5$BOLD} + {\tt PHOENIX} &\teff=6500 K, $\log(g)=4.0$, 12 weighted structures &5.409$\pm$0.011   &5.393$\pm$0.005  &5.405$\pm$0.006  \\
``D'' {\tt ATLAS 12}   &\teff=6530 K, $\log(g)=3.95$, $\alpha=1.25$, no overshooting   	&5.426$\pm$0.011   &5.404$\pm$0.005  &5.404$\pm$0.006  \\
``E'' {\tt ATLAS 12}   &\teff=6530 K, $\log(g)=3.95$, $\alpha=1.25$, 100\% overshooting &5.368$\pm$0.011   &5.383$\pm$0.005  &5.394$\pm$0.006  \\
``F'' {\tt ATLAS 12}   &\teff=6530 K, $\log(g)=3.95$, $\alpha=1.25$, 50\% overshooting	&5.400$\pm$0.011   &5.398$\pm$0.005  &5.402$\pm$0.006  \\
``U''  Uniform Disk    &$V^2 = \left| 2J_1(\pi\theta_{\rm UD}B/\overline{\lambda})/(\pi\theta_{\rm UD}B/\overline{\lambda})\right|^2$   &5.044$\pm$0.010   &5.208$\pm$0.005  &5.302$\pm$0.005  \\
\enddata
\label{tab:AD_fits}
\end{deluxetable}

\begin{deluxetable}{llllrr}
\tablecolumns{6} 
\tabletypesize{\footnotesize}
\tablecaption{Synthetic Str\"{o}mgren Indices}
\tablewidth{0pt}
\tablehead{
\colhead{Model}	    
&\colhead{\teff (K)}	    
&\colhead{$\log(g)$}	    
&\colhead{Additional Parameters}	    
&\colhead{$(b-y)$\tablenotemark{a}}
&\colhead{$c_1$\tablenotemark{a}}\\
\colhead{}	    
&\colhead{}	    
&\colhead{}	    
&\colhead{}	    
&\colhead{}
&\colhead{}
}
\startdata              
{\tt ATLAS 12} ``E''   &6530  &3.95 &$\alpha=1.25$, overshooting        &0.321  &0.378  \\
{\tt ATLAS 12} ``F''   &6530  &3.95 &$\alpha=1.25$, 50\% overshooting 	&0.311  &0.402  \\
{\tt ATLAS 12} ``D''   &6530  &3.95 &$\alpha=1.25$, no overshooting   	&0.302  &0.437  \\
{\tt PHOENIX}  ``A''   &6530  &3.95 &$\alpha=1.25$                   	&0.292  &0.519  \\
{\tt PHOENIX}  ``B''   &6530  &3.95 &$\alpha=0.50$                      &0.282  &0.547  \\
Composite SED A       &6466  &3.95 &$\beta$ = 0.60 linear limb darkening     &0.293   &0.539  \\
Composite SED B       &6531  &3.95 &$\beta$ = 0.50 linear limb darkening     &0.283   &0.553  \\
\enddata 
\tablenotetext{a}{Interstellar reddening towards Procyon is assumed to be negligible: $(b-y) = (b-y)_0$, $c_1=c_0$.
Observed values: $(b-y) = 0.272$ \cite[]{cb70, nord04} and  $c_1 = 0.532$ \citep[]{cb70}.}
\label{tab:colors}
\end{deluxetable}

\begin{deluxetable}{llcl}
\tablecolumns{4} 
\tabletypesize{\footnotesize}
\tablecaption{Fundamental Parameters for Procyon A} 
\tablewidth{0pt}
\tablehead{
\colhead{Parameter}
&\colhead{Symbol}
&\colhead{Value}
&\colhead{Reference}
}
\startdata              
Limb-darkened angular diameter (mas)  &\adld           &5.404$\pm$0.031                 &Table~\ref{tab:AD_fits}, \S~\ref{discussion}\\
Bolometric flux (erg cm$^{-2}$ s$^{-1}$)  &$\mathcal{F}$   &$(17.86\pm0.89)\times 10^{-6}$ &\S~\ref{teff} \\
Effective temperature (K)             &\teff           &6543$\pm$84                    &\adld\ and $\mathcal{F}$\\
Parallax  (mas)                       &$\pi$           &285.93$\pm$0.88                &\cite{hipparcos}  \\
Radius (\rsun)                        &$R$             &2.031$\pm$0.013                &\adld\ and $\pi$ \\
Mass   (\msun)                        &$M$             &1.42$\pm$0.04                  &\cite{girard2000} \\
Surface gravity (cm s$^{-2}$)         &$\log(g)$       &3.975$\pm$0.013                &$R$ and $M$ \\
\enddata 
\label{tab:fund_params}
\end{deluxetable}

\end{document}